\documentclass[aps,prd,superscriptaddress,twocolumn,nofootinbib,showpacs]{revtex4}

\usepackage{graphicx}
\usepackage{dcolumn}
\usepackage{bm}
\usepackage{epsfig}
\usepackage{epstopdf} 
\usepackage{multirow}
\usepackage{subfigure}
\usepackage{lineno}
\usepackage{rotating}
\def\pt{\ensuremath{p_{\mathrm{T}}}} 
\def\mc{Monte Carlo} 
\def\sm{Standard Model} 
\def\fr{factorization and renomalization} 
\def\vbf{weak-boson-fusion}
\def\me{matrix-element}
\def\hjjph{H($\to b\bar{b}$)+2jets+$\gamma$}
\def\bbjjph{$b\bar{b}$+2partons+$\gamma$}
\def\bbjph{$b\bar{b}$+1parton+$\gamma$}

\def\bjet{b-jet}
\def\cjet{c-jet}
\def\ljet{light-jet}
\def\pp{$pp$}

\begin{document}

\newcommand{\spartyjet}{{\sc spartyjet}}
\newcommand{\antikt}{{\sc anti-kt}}
\newcommand{\siscone}{{\sc siscone}}
\newcommand{\pythia}{{\sc Pythia}}
\newcommand{\herwig}{{\sc Herwig}}
\newcommand{\alpgen}{{\sc Alpgen}}


\title{Prospects for Observing the Standard Model Higgs Boson Decaying into $b\bar b$ Final States Produced in Weak Boson Fusion with an Associated Photon at the LHC}


\author{D. M. Asner}
\author{M. Cunningham}
\author{S. Dejong}
\author{K. Randrianarivony}
\affiliation{Carleton University, Ottawa, Ontario, Canada K1S 5B6}

\author{C. Santamarina}
\author{M. Schram}
\affiliation{McGill University, Montreal, Quebec, Canada H3A 2TS}

\noaffiliation

\parindent=0in



\date{\today}

\begin{abstract} 
One of the primary goals of the Large Hadron Collider is to understand the electroweak symmetry breaking mechanism.
In the Standard Model, electroweak symmetry breaking is described by the Higgs mechanism which includes a scalar Higgs boson.
Electroweak measurements constrain the Standard Model Higgs boson mass to be in the 114.4 to 157 GeV/c$^{2}$ range.
Within this mass window, the Higgs predominantly decays into two b-quarks.
As such, we investigate the prospect of observing the Standard Model Higgs decaying to $b\bar{b}$ produced in \vbf\ with an associated central photon.
An isolated, high \pt, central photon trigger is expected to be available at the ATLAS and CMS experiments.
In this study, we investigated the effects originating from showering, hadronization, the underlying event model, and jet performance including \bjet\ calibration on the sensitivity of this channel.
We found that the choice of \mc\ and \mc\ tune has a large effect on the efficacy of the central jet veto and consequently the signal significance.
A signal significance of about 1.86 can be achieved for $m_{h}=115$ GeV/c$^{2}$ with 100 fb$^{-1}$ of integrated luminosity which correspond to one year at design luminosity at 14 TeV \pp\ collisions.
\end{abstract}
\maketitle


\section{Introduction}\label{S:intro}
\begin{figure}
\begin{center}
\includegraphics[width =0.15\textwidth]{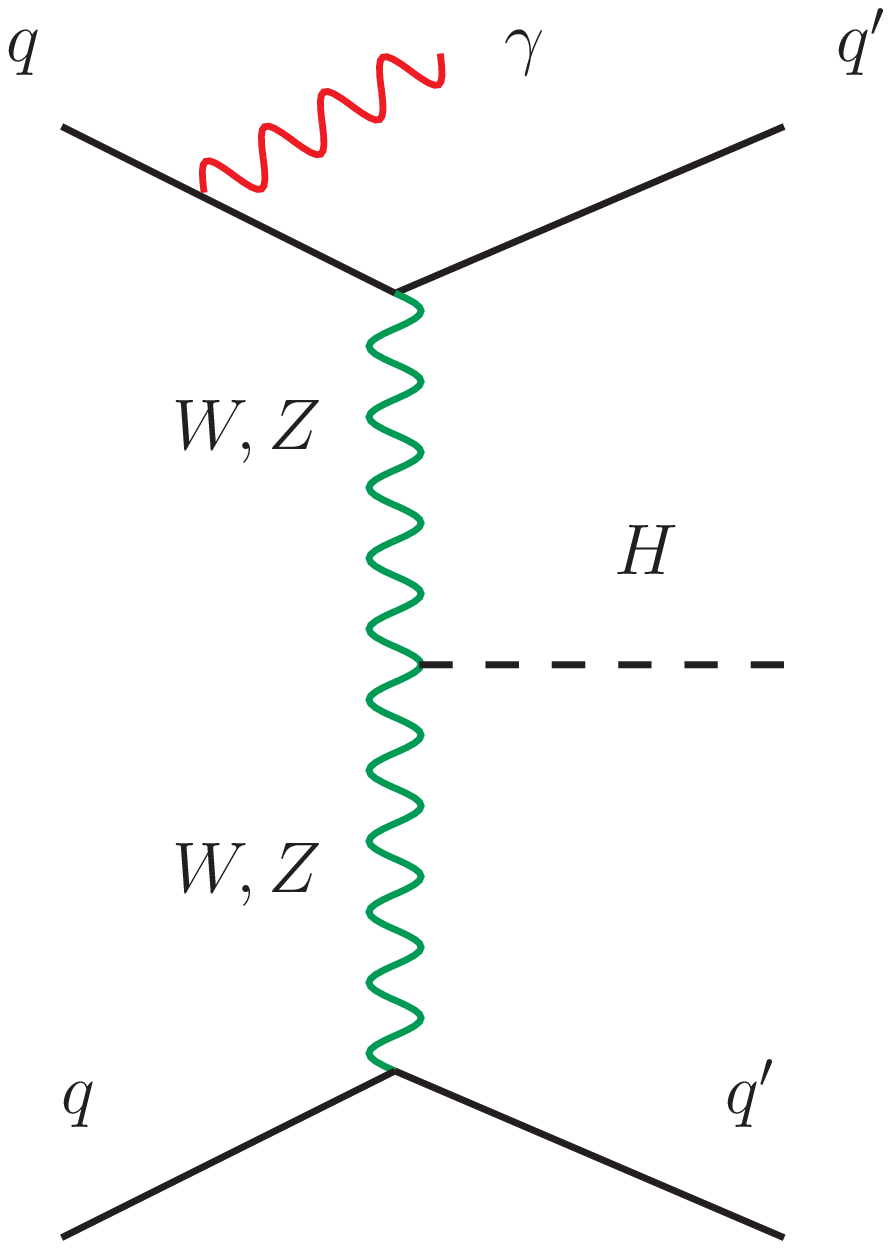} 
\includegraphics[width =0.15\textwidth]{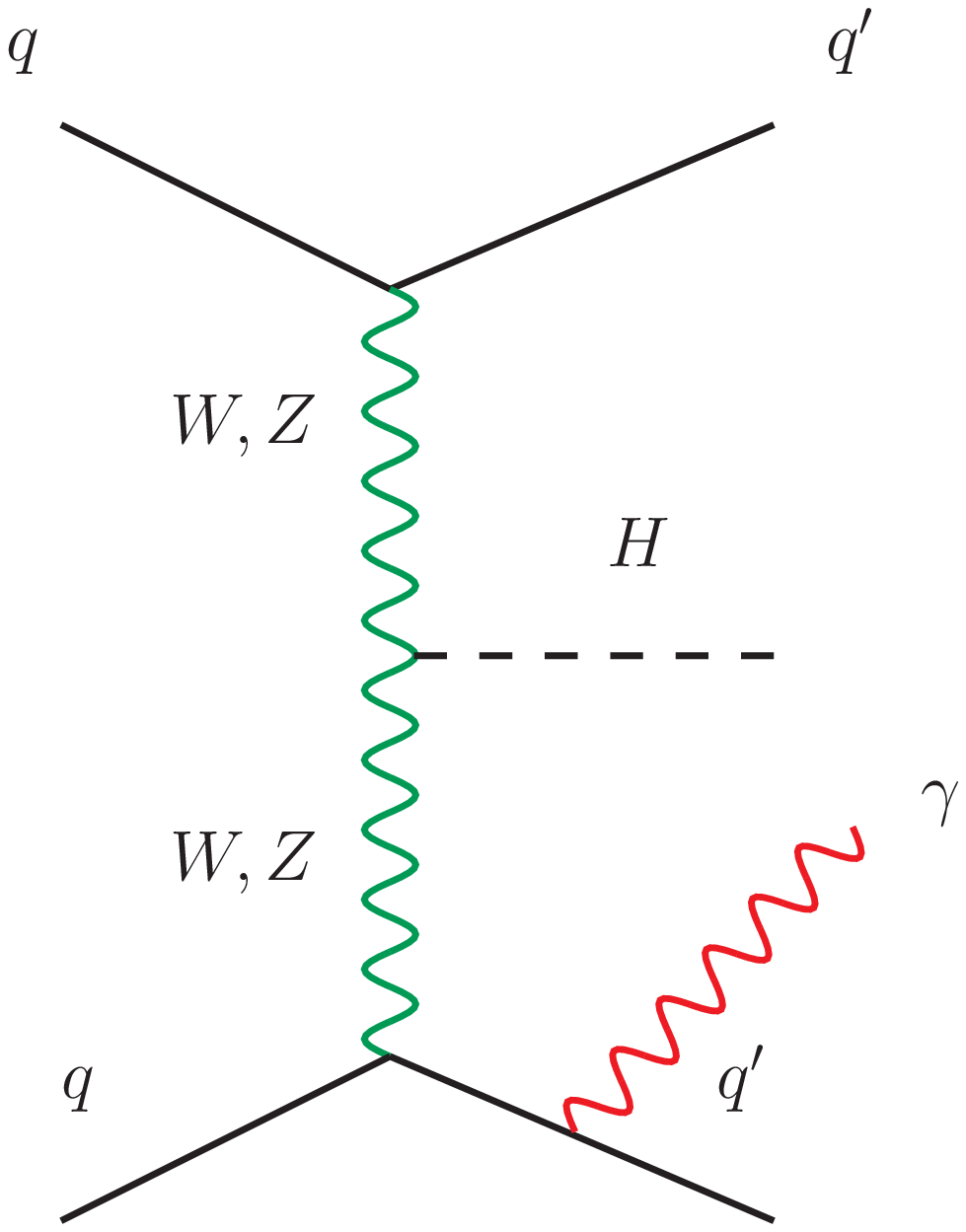}
\includegraphics[width =0.15\textwidth]{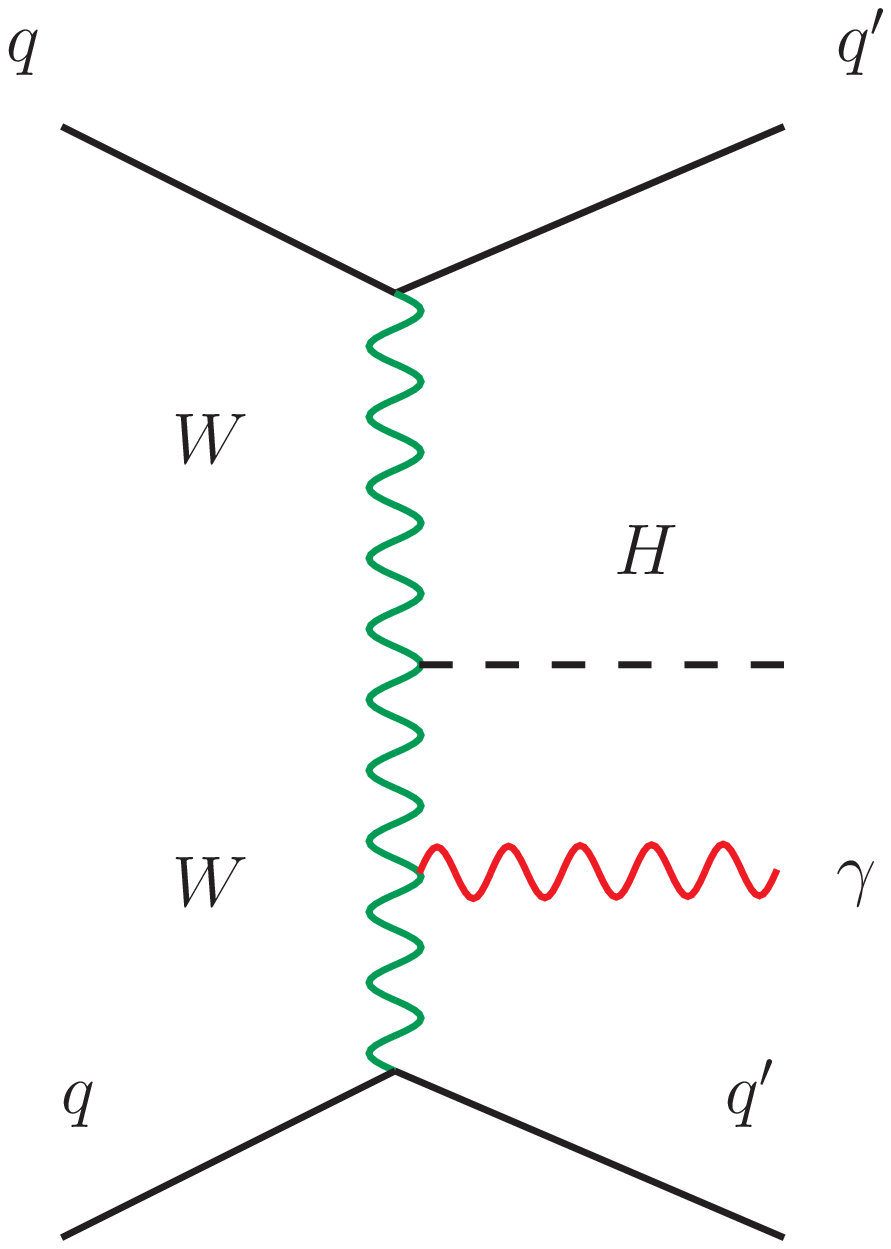} 
\caption{Tree-level t-channel Feynman diagrams for Higgs boson production in the process $pp\to H\gamma jj$. 
Here, $q$ and $q^\prime$ represent quarks ($u$, $d$, $s$, $c$) while $q$ and $q^\prime$ are the same when a $Z$ boson is exchanged.}
\label{F:sig_Feyn}
\end{center}
\end{figure}

One of the primary goals of the Large Hadron Collider (LHC) is to investigate the electroweak symmetry breaking mechanism which is explained in the  \sm\ (SM)~\cite{StandardModelRef}, by the Higgs mechanism~\cite{HiggsRef}.
Although the Higgs boson has yet to be discovered, there are several theoretical and experimental constraints on the \sm\, Higgs mass.
The lower bound on the Higgs mass of 114.4 GeV/c$^{2}$ at 95$\%$ Confidence Level (C.L.) is constrained by direct measurements at LEP~\cite{LEP_FINAL}.
Additionally, the upper bound is constrained indirectly by several global fits to electroweak measurements.
A global fit of the electroweak data as a function of Higgs mass reported by LEP Electroweak Working Group on the favors a Higgs mass centered at $87^{+35}_{-26}$ GeV/c$^{2}$ and a 95$\%$ one-sided C.L. on the upper limit at 157 GeV/c$^{2}$.
Similarly, the GFitter results using a global fit which includes the constraints from the direct Higgs boson searches yields an upper limit on the Higgs mass of 153 GeV/c$^{2}$  at 95$\%$ C.L.~\cite{GFitterOct09}.
Within the favored Higgs mass region the range excluded at 95$\%$ C.L. for a SM Higgs is 163 $< m_{H} <$ 166 GeV/c$^{2}$ as determined by the Tevatron~\cite{TevNov19_2009}. \\ \\
There are several channels, such as $H\rightarrow \gamma \gamma$,  $H\rightarrow \tau \tau$, and $H\rightarrow Z^{*} Z$, which are expected to discover a light SM Higgs at the LHC with approximately 30 fb$^{-1}$ of 14 TeV \pp\ data~\cite{CMS_2006,CSC_2009}.
However, additional studies will be required to confirm if the new resonance is indeed the SM Higgs.\\ \\
In this paper, we study the prospect for observing the process $H\rightarrow b\bar{b}$ which will be instrumental in determining spin, CP, gauge coupling, and Yukawa coupling of the Higgs candidates.
Distinguishing $H\rightarrow b\bar{b}$ from the large QCD background is the main challenge for this analysis. 
At the LHC, the largest production cross section for Standard Model Higgs boson is gluon fusion ($gg \to H$); the next largest is Weak-Boson-Fusion (WBF).
The distinct kinematical and QCD properties of the WBF production, described in Section \ref{S:WBF}, provide discrimination from QCD processes. 
We resolve the challenge of triggering on a four-jet final state by requiring an associated photon.
Specifically, we investigate the sensitivity to a light SM Higgs Boson produced in association with a photon in WBF production, as shown in Figure~\ref{F:sig_Feyn}, in 14 TeV \pp\ collisions at the LHC.
The photon provides a simple unprescaled trigger for the four-jet + photon final state as originally proposed by E. Gabrielli, {\em et al.}~\cite{VBFGamma07}.
\\ \\
The \mc\, samples generated for this study are described in Section \ref{S:Evg}.
In Section \ref{S:ana}, the selection criteria used in the analysis are presented.
The event selection is composed of photon trigger and identification (Section \ref{S:Photon}), jet performance studies (Section \ref{S:JET_PERF}), \vbf\, jet identification (Section \ref{S:VBF_ID}), \bjet \,identification and light jet fake rates (Section \ref{S:B_ID}), \bjet\ calibration (Section \ref{S:bjet_calib}), and central jet veto (Section \ref{S:CJV}).
In Section \ref{S:sys_uncertainty}, the systematic uncertainties associated with the \mc\, generator cross-sections and the choice of showering, hadronization, and the Underlying Event (UE) model are presented.
Finally, the results and summary of this analysis are discussed in Sections \ref{S:results} and \ref{S:summary} respectively.
\section{Weak Boson Fusion}\label{S:WBF}
The WBF process consists of weak gauge bosons radiated from the quark partons fusing to become a Higgs boson. 
The quarks scatter with sufficiently high \pt\ to be detected in the forward hadronic calorimeters of ATLAS and CMS. 
The Higgs boson is produced centrally as are the two b-jets. 
The kinematics of this four-jet topology, one WBF forward, one backward and two central jets, is exploited to suppress the four-jet background from QCD processes. 
Additionally, jet activity in signal events tends to be forward of the WBF jets. 
QCD radiation is at small angles with respect to the WBF quarks with no color connection between the scattered quarks. 
In contrast, QCD production involves color charge exchange between the scattered quarks. 
Consequently, the QCD radiation takes place over large angles and populates the central detector region. 
The optimization of the selection criteria is described in Section \ref{S:ana}.
\section{Event Samples and Simulation}\label{S:Evg}
In this paper, the signal and background Monte Carlo events were produced in two stages using \alpgen~\cite{Alpgen_2003} and \pythia~\cite{Pythia64_2006}. 
The \alpgen\,generator was used in the first stage to produce the parton four-vectors and to determine the leading order cross-sections. 
The hard partonic interaction used at the generator stage was evaluated using the CTEQ6L1 structure functions~\cite{PDF_2002} for the colliding protons. 
Several kinematic cuts were applied on the signal and background samples in the analysis and are listed in Table~\ref{T:AlpgenCutsMLM}.
Within this paper $\Delta R$ is defined as  $\Delta R_{kl} = \sqrt{\Delta \eta_{kl}^{2} + \Delta \phi_{kl}^{2}}$, $\Delta \eta_{kl}=\eta_{k}-\eta_{l}$, $\Delta \phi_{kl}=\phi_{k}-\phi_{l}$ of the $k^{th}$ and $l^{th}$ parton, where $\eta$ is the pseudo-rapidity and $\phi$ is the azimuthal angle.
\begin{table}[h]
\begin{center}
\begin{tabular}{|cccc|}\hline \hline
$\pt(j)>15$ GeV/c & $|\eta_{j}|<5.5$ & $\Delta R_{j j}>0.7$ & \\
$\pt(\gamma)>15$ GeV/c &  $|\eta_{\gamma}|<3.0$ &  $\Delta R_{\gamma j}>0.7$ &  $\Delta R_{\gamma b}>0.7$ \\
$\pt(b)>15$ GeV/c& $|\eta_{b}|<5.5$  & $\Delta R_{b j}>0.7$ & $\Delta R_{b\bar{b}}>0.7$ \\
\hline \hline
\end{tabular}
\caption{\label{T:AlpgenCutsMLM} \alpgen\ generator level kinematic cuts applied on signal and background.}
\end{center}
\end{table}
To provide inclusive background samples without double counting, a jet-parton matching scheme referred to as the MLM prescription~\cite{MLM_2006} was applied. 
In this paper, the jet-parton ($jp$) matching efficiency is referred to as the MLM efficiency.  
The matching requirements used for this analysis were  $\pt(j)>17.5$ GeV/c, $\Delta R_{jp}<0.7$, and $|\eta_{j}|<6.0$.\\ \\
\pythia\, was used for the second stage of the \mc\ production to perform the heavy particle decays, including the Higgs decay, showering, hadronization, and the UE model.
The resulting cross-sections are provided in Sections \ref{S:HiggsEvg} and \ref{S:BkgEvg}.
\subsection{Higgs Signal}\label{S:HiggsEvg}
The Feynman diagrams for the signal process are shown in Figure~\ref{F:sig_Feyn}.
Three mass points were used to scan the range in which the Higgs decaying to two b-quarks is dominant.
The nominal parameterization of the  \fr\,
scales was set to $\mu_{\rm{F}}^{2}=\mu_{\rm{R}}^{2}=m^{2}_{h}+
p^{2}_{\rm{T}}(\gamma)+\sum p^{2}_{\rm{T}}(j)$, where
$p_{\rm{T}}(\gamma)$ and $p_{\rm{T}}(j)$ are the transverse
momentum of the photon and of the jets, respectively. 
The sum over the transverse momentum includes all final state jets.
Table \ref{T:HiggsAlpgenGen} shows the cross-section and branching fraction obtained for each Higgs mass point.
The Higgs branching fraction was evaluated with the program HDECAY~\cite{Djouadi97}.
\begin{table}[h]
\centering
\begin{tabular}{|c|c|c|c|}\hline \hline
$m_{h}$ [GeV/c$^{2}$] & 115 & 125 & 135\\ \hline
$\sigma (H\gamma jj)$ [fb] & 69.7 & 67.0 & 52.8 \\
$\mathrm{BR}(H\rightarrow b\bar{b})$  & 0.73 & 0.61 & 0.43 \\ \hline
$\sigma \times \mathrm{BR} $ [fb] & 50.9 & 40.9 & 22.7 \\
\hline \hline
\end{tabular}
\caption{\label{T:HiggsAlpgenGen} Cross sections for the $H\gamma jj$ signal  for 14~TeV \pp\ collisions.
Additionally, the Higgs boson branching fraction to $b\bar{b}$ using HDECAY~\cite{Djouadi97}.}
\end{table}
\subsection{Backgrounds}\label{S:BkgEvg}
\begin{figure}[hbt]
\begin{center}
\includegraphics[width =0.15\textwidth]{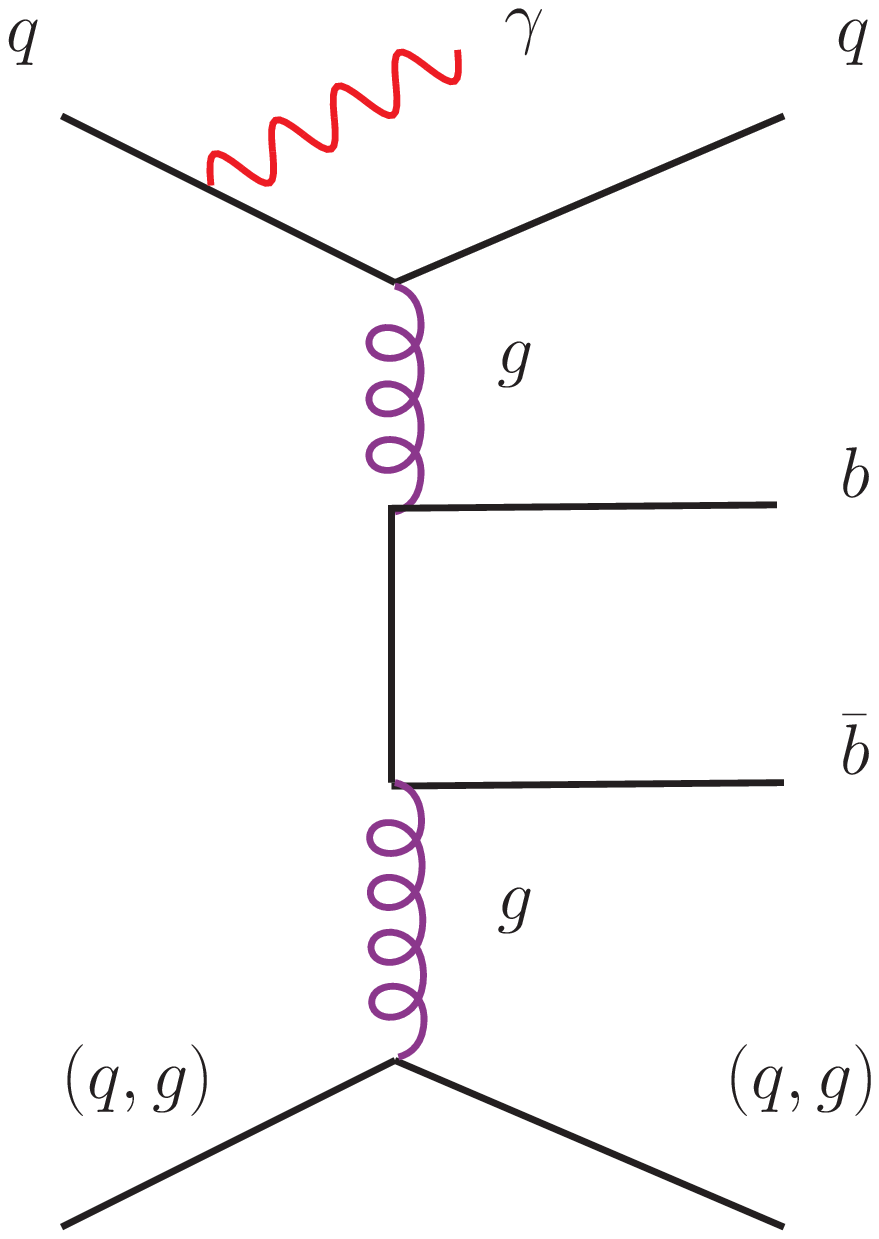}
\includegraphics[width =0.15\textwidth]{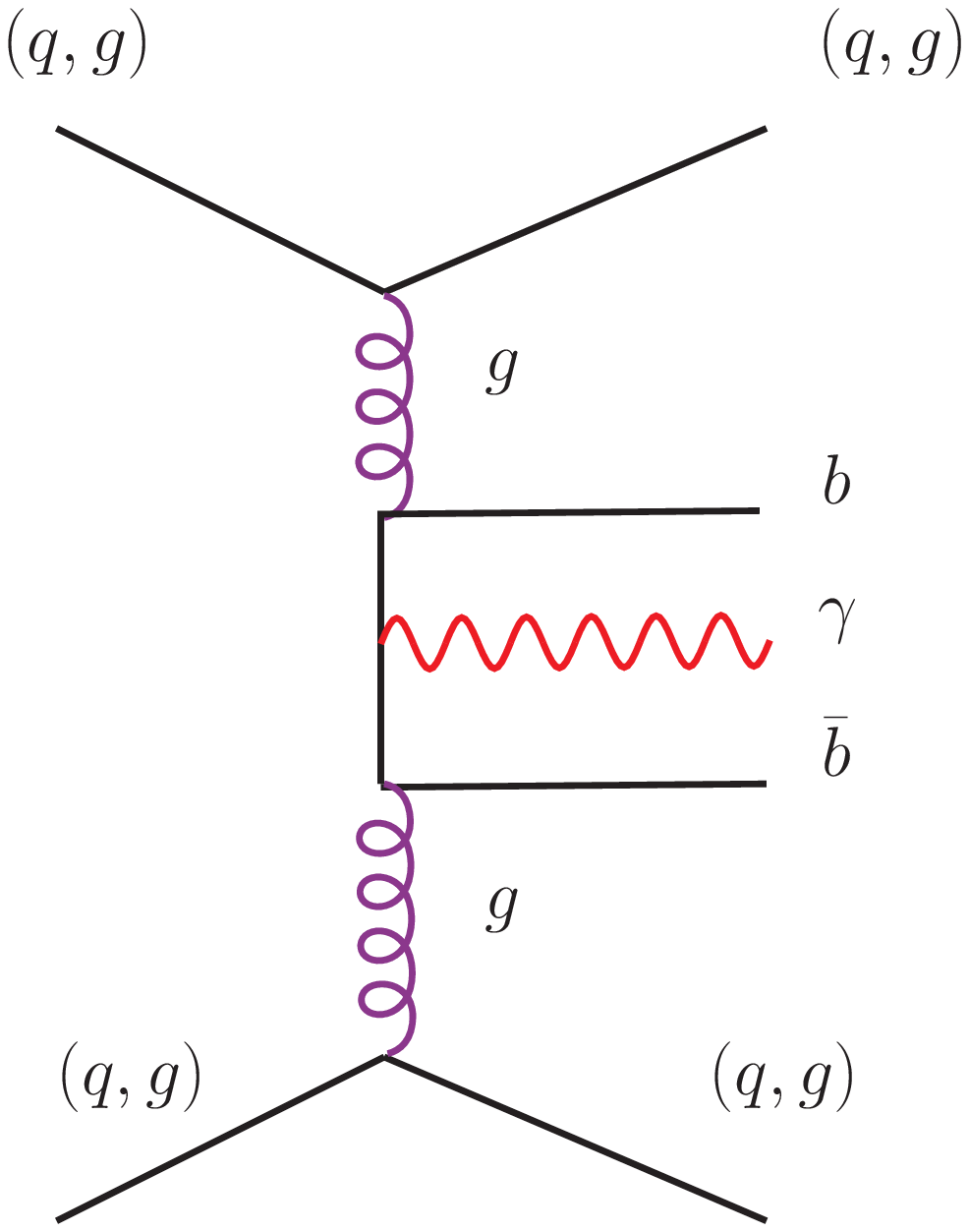}\\ 
\includegraphics[width =0.15\textwidth]{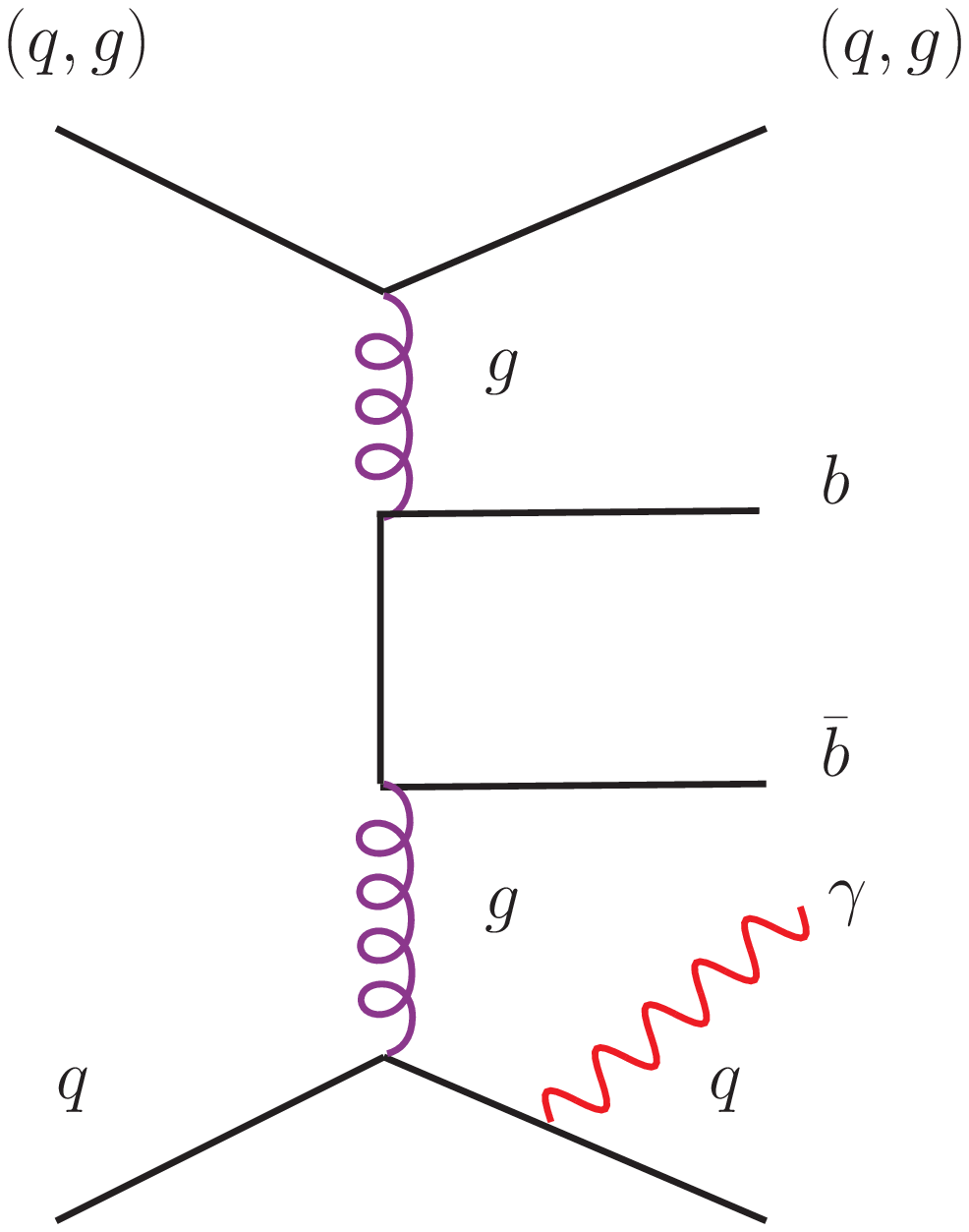}
\includegraphics[width =0.15\textwidth]{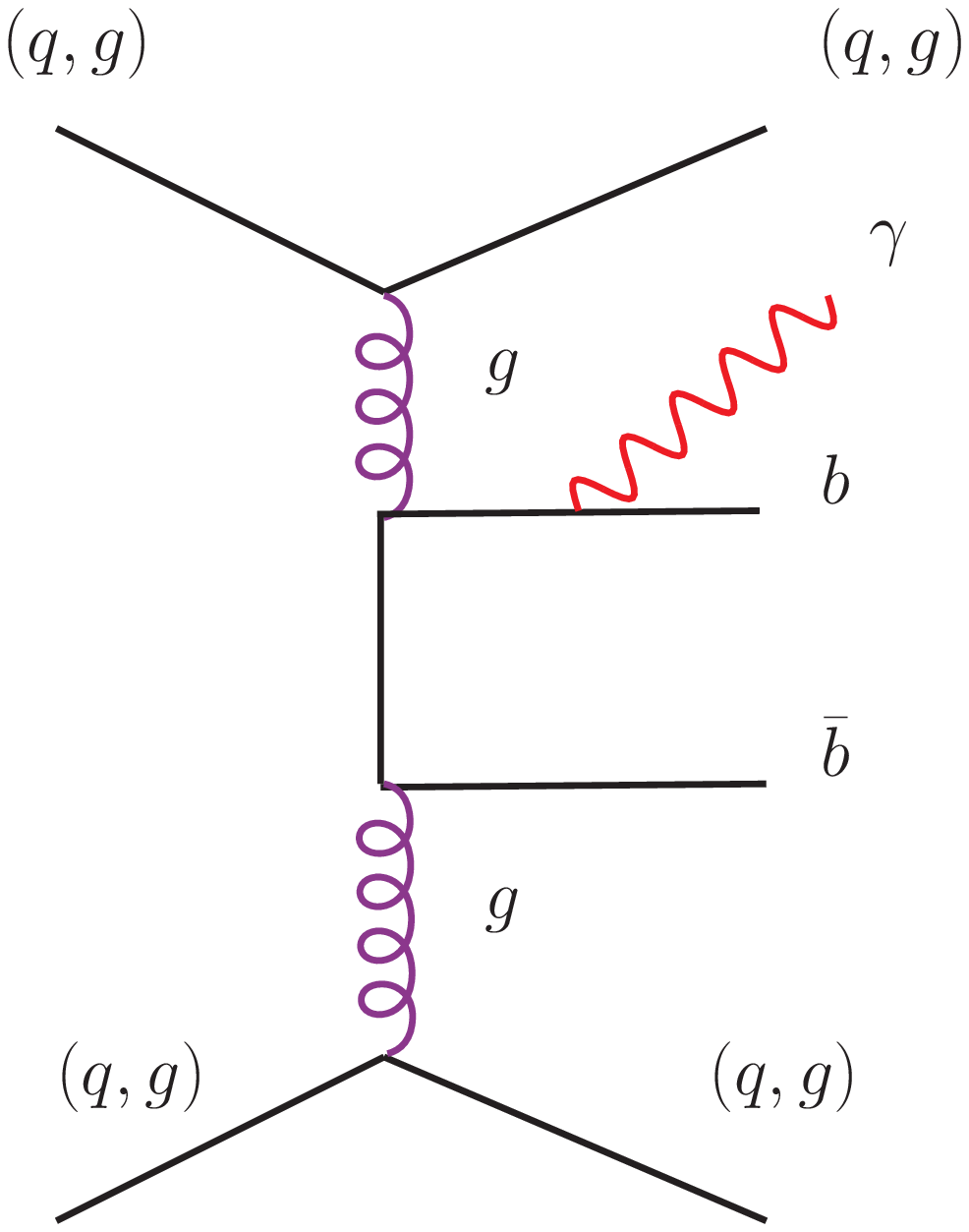}
\caption{Representative classes of Feynman diagrams contributing, at parton level, to the 
background process $pp \to bbjj\gamma$. Here $q$ and $g$ represent the quarks ($u$, $d$, $s$, $c$) and gluon, respectively.}
\label{F:bkgd_Feyn}
\end{center}
\end{figure}

The event topology consists of four jets and an associated photon.
Consequently, the largest backgrounds originate from QCD processes with radiated photon(s).\\ \\
The primary background is the $b\bar{b}$+2partons+$\gamma$ process, shown in Figure \ref{F:bkgd_Feyn}, which has the same final state particles, that is, one photon, two \bjet s, and 2 additional jets.
Additionally, the large background cross-section relative to that of the signal makes the combinatorial invariant mass of the two b-quarks very challenging to suppress.
We also considered the $b\bar{b}+$1parton$+\gamma$ in which a secondary jet can originate from showering, hadronization, or the UE model.
The $b\bar{b}$+$n$partons+$\gamma$ were produced such that the 1 parton sample is exclusive and the 2 partons sample is inclusive.
These samples were produced using the \fr\, scales $\mu_{\rm{F}}^{2}=\mu_{\rm{R}}^{2}=\sum m^{2}_{\rm{T}}$, where the sum includes all final state partons, $m_{\rm{T}}$ is the transverse mass defined as $m^{2}_{\rm{T}}=m^{2}+ p^{2}_{\rm{T}}$, and $m$ is the mass of the parton. \\ \\
The Z($\rightarrow q\bar{q}$)+$n$parton+$\gamma$ sample was considered because the high side of the Z-boson mass peak can potentially contaminate the signal region.
The samples were produced such that all quark flavors were included except for the top quark. 
For similar reasons as the $b\bar{b}$+$n$partons$+\gamma$ background, we produced a one parton exclusive sample and a 2 partons inclusive sample.
The  \fr\,  scales were set to $\mu_{\rm{F}}^{2}=\mu_{\rm{R}}^{2}=m^{2}_{Z}+p^{2}_{\rm{T}}(\gamma)+\sum p^{2}_{\rm{T}}(j)$.\\ \\
In addition to background processes which have legitimate b-quarks, we also considered light quark and gluon QCD processes which fake reconstructed \bjet s.
We only considered processes with large cross-sections because of the low probability of two light jets faking two b-jets.
Consequently, we produced $n$parton+$\gamma$ samples,  with an exclusive 3 partons and inclusive 4 partons sample. 
We neglected the $n$parton+$2\gamma$ processes which are suppressed by an additional factor of $\alpha_{em}$. \\ \\ 
The \alpgen\, cross-sections, the MLM efficiencies, and effective cross-sections for the aforementioned background samples are presented in Table \ref{t:bgd_xs_effic}.\\ \\
\begin{table}[htdp]
\begin{center}
\begin{tabular}{|c|c|c|c|}
\hline \hline
Parton \# & $\sigma$ [pb] @14~TeV & MLM $\epsilon$ [\%] & $\sigma^{\prime}=\sigma\times$(MLM $\epsilon)$ [pb]  \\
\hline
\multicolumn{4}{|c|}{\multirow{2}{*}{$b\bar{b}$+$n$parton+$\gamma$}} \\
\multicolumn{4}{|c|}{}  \\  \hline
1 & 1088 & 37.9  & 452 \\
2 &  658  & 35.7  & 235 \\
\hline
\multicolumn{4}{|c|}{\multirow{2}{*}{$n$parton+$\gamma$}} \\
\multicolumn{4}{|c|}{}  \\  \hline
3 & 45789 & 20.5 & 9387 \\
4 & 17595 & 17.8 & 3130 \\
\hline
\multicolumn{4}{|c|}{\multirow{2}{*}{$Z(\to q \bar{q})$+$n$parton+$\gamma$}} \\
\multicolumn{4}{|c|}{}  \\  \hline
1 & 27.1 & 40.6  & 11.0 \\
2 & 18.2 & 44.5 &  8.10 \\
\hline \hline
\end{tabular}
\caption{\label{t:bgd_xs_effic} Cross-section, MLM efficiency, and the effective cross-section for background processes as determined using \alpgen/\pythia\ for 14 TeV \pp\ collisions.}
\end{center}
\end{table}

\section{Event Selection}\label{S:ana}
In this section, we present the analysis strategy to measure the Higgs boson using the \vbf\ production mechanism with an associated photon production.
A Higgs event in this process is characterized by two geometrically well separated jets, a prompt photon, and the Higgs decay products.
In Section \ref{S:Photon}, our assumptions on photon trigger and photon identification are summarized.
The jet performance is presented in Section~\ref{S:JET_PERF}.
Identification strategies for the  \vbf\ jets are described in Section~\ref{S:VBF_ID}.
In this analysis, we required the Higgs to decay into two b-quarks.
Consequently,  Section \ref{S:B_ID} summarizes the \bjet\, identification and fake rate estimation from \cjet s and light jets.
We discuss the \bjet\ calibration in Section \ref{S:bjet_calib}.
In Section \ref{S:CJV}, we investigate the possibility of applying a veto on additional central jets given the present theoretical models for the UE.\\ \\
In order to maximize the signal significance, we performed a ranked optimization using signal over square root of background ($S/\sqrt{B}$) as the figure-of-merit.
For the signal, we used an independent sample with a Higgs mass of 115 GeV/c$^{2}$ and used the \bbjjph\  sample to estimate the full background.
For the \bbjjph\  sample, we used events in sideband region invariant mass of the two \bjet's, $m_{bb}$, sideband which was defined as 100 GeV/c$^{2}$ $<m_{bb}$ and $m_{bb}>$130 GeV/c$^{2}$ in order to avoid any bias.
\subsection{Trigger and Photon Identification}\label{S:Photon}
\begin{figure}[hbt]
\begin{center}
\includegraphics[width =0.45\textwidth]{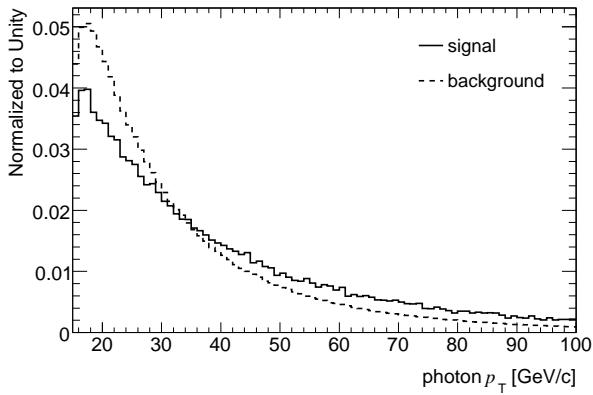}
\caption{Transverse momentum distribution of the associated photon for a 115 GeV/c$^{2}$ mass Higgs at 14 TeV collision energies.}
\label{F:gamma_pt}
\end{center}
\end{figure}

An isolated tight 25 GeV/c transverse momentum photon trigger is assumed to be available at the LHC experiments.
In addition, the trigger turn-on curve is expected to plateau very quickly requiring a modest increase in the transverse momentum of reconstructed photons. 
Consequently, for this analysis we required an isolated 30 GeV/c transverse momentum photon within $|\eta|<2.5$, corresponding to the coverage of the electromagnetic calorimeters in ATLAS and CMS~\cite{CMS_2006,CSC_2009}.
The photon is identified by matching all candidates to the matrix-element photon.
Once the proper photon is identified, we removed any overlapping particles or jets within a $\Delta R$ of 0.7. 
The \pt\ spectrum of the photons is shown in Figure \ref{F:gamma_pt}.

\subsection{Jet Identification and Performance}\label{S:JET_PERF}
In addition to the central photon, which provides a viable trigger option and reduces the cross section of the backgrounds, the only additional objects used for this analysis are jets. 
Consequently, the aim of this section is to study the effect of the jet performance on the potential to observe the \hjjph\ channel.\\ \\
The goal of jet reconstruction is to reproduce the 4-momentum of the original parton.
However, the showering, hadronization, particle decays, and detector performance complicates this goal.
In this section, we investigate the performance of jet algorithms at the hadronization level.
In order to account for some detector effects, we excluded muons and neutrinos from the jet constituent input list as these particles are not measured by the calorimeter.
Additionally, we excluded charged particles with $\pt < 400$ MeV/c and neutral particles with $\pt < 200$ MeV/c.
For this study, we evaluated the performance of the \antikt~\cite{AntiKt_2008} and \siscone~\cite{SisCone2007} algorithms as provided in the \spartyjet~\cite{SpartyJet2008} package.
For \siscone\, we considered a cone radius of 0.4 and 0.7 (labeled as \siscone7 and \siscone4) and the aggregation distance of 0.4 and 0.6 for \antikt\ (labeled as \antikt4 and \antikt6). \\ \\
The results presented in this section were obtained using the signal \mc\ sample with a 115 GeV/c$^{2}$ Higgs mass\footnote{The \bbjjph\ \mc\ samples yielded similar results.}.
We studied two performance requirements in order to compare the various jet algorithms and settings.
These requirements are listed below.
\begin{enumerate}
\item  The efficiency to identify each individual parton.
\item  The ability to properly reconstruct the transverse momentum (linearity and resolution).
\end{enumerate}
\subsubsection{Efficiency}
\begin{figure}
\begin{center}
\includegraphics[width=0.45\textwidth]{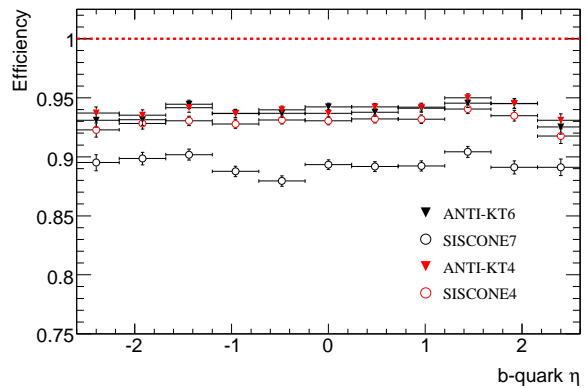}
\caption{\label{efficiency_b_eta} Efficiency for b-quark reconstruction as a jet as a function of $\eta$ using a 115 GeV/c$^{2}$ Higgs mass sample.
The small asymmetry in $\eta$ is due to the seeding/sorting of jets.}
\end{center}
\end{figure}

\begin{figure}
\begin{center}
\includegraphics[width=0.45\textwidth]{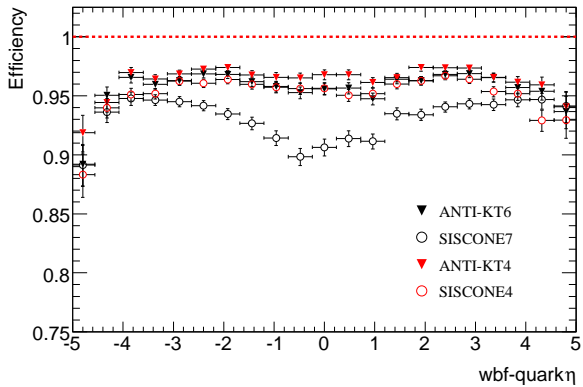}
\caption{\label{efficiency_vbf_eta} Efficiency for WBF-quark reconstruction as a jet as a function of $\eta$ using a 115 GeV/c$^{2}$ Higgs mass sample.
The small asymmetry in $\eta$ is due to the seeding/sorting of jets.}
\end{center}
\end{figure}

\begin{figure}
\begin{center}
\includegraphics[width=0.45\textwidth]{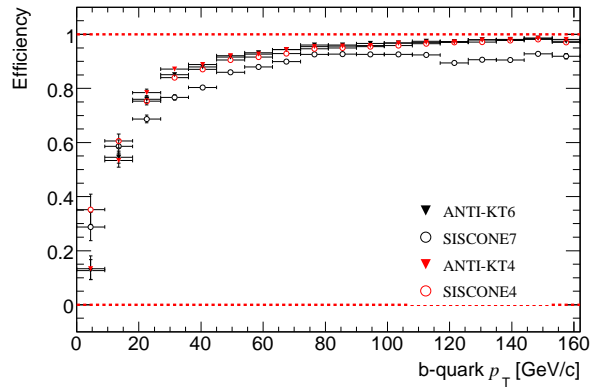}
\caption{\label{efficiency_b_pt}Efficiency for b-quark reconstruction as a jet as a function of the b-quark \pt\ using a 115 GeV/c$^{2}$ Higgs mass sample. Only events with $|\eta|<2.5$ were considered.}
\end{center}
\end{figure}

\begin{figure}
\begin{center}
\includegraphics[width=0.45\textwidth]{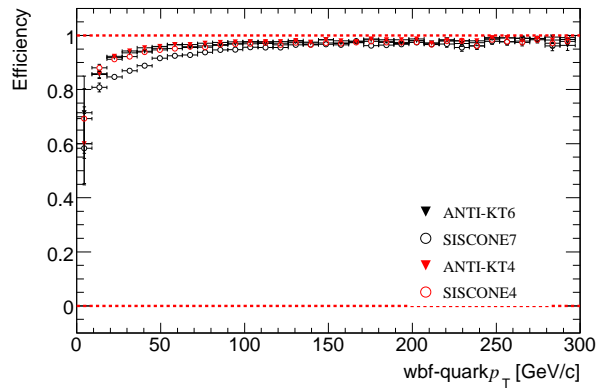}
\caption{\label{efficiency_vbf_pt}Efficiency for WBF-quark reconstruction as a jet as a function of the WBF-quark \pt\ using a 115 GeV/c$^{2}$ Higgs mass sample.. }
\end{center}
\end{figure}

\begin{figure}
\begin{center}
\includegraphics[width=0.45\textwidth]{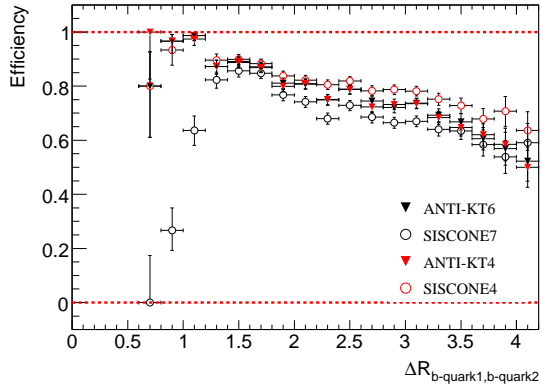}
\caption{\label{efficiency_dr_b} Efficiency to reconstruct two b-quarks as two jets as a function of the $\Delta R$ between the two quarks.}
\end{center}
\end{figure}

\begin{figure}
\begin{center}
\includegraphics[width=0.45\textwidth]{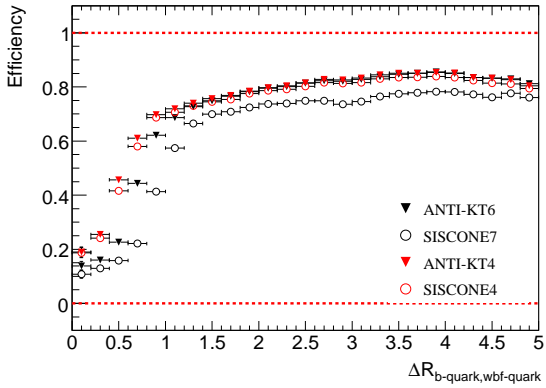}
\caption{\label{efficiency_dr_vbf} Efficiency for a WBF-quark b-quark jet reconstruction as a function of the $\Delta R$ between the two quarks.}
\end{center}
\end{figure}

We considered four factors that can contribute to the jet reconstruction inefficiency 
\begin{enumerate}
\item Parton constituents outside of the detector acceptance window.
\item Low transverse momentum partons leading to particles with insufficient energy after hadronization.
\item Neutrinos - which are not detected.
\item Small angular separation between outgoing partons resulting in a single jet.
\end{enumerate}
To study the jet inefficiency for a single parton, we selected isolated quarks by requiring that no additional jets are within $|\Delta \phi|<1.4$.
Additionally, we defined a matched jet as the jet closest to the outgoing parton within $\Delta R < 0.4$.
The efficiencies as a function of $\eta$ for \bjet s and WBF jets are shown in Figures \ref{efficiency_b_eta} and \ref{efficiency_vbf_eta}, respectively.
The effect of the detector acceptance window near $\eta=5$ is visible in Figure~\ref{efficiency_vbf_eta}.
The effect of inefficiency as a function of the initial parton transverse momentum can be seen in Figures~\ref{efficiency_b_pt} and~\ref{efficiency_vbf_pt}.
Here we required the b-quark and WBF quarks to be within $|\eta|<2.5$ and $|\eta|<4.5$, respectively.
We restrict the \bjet $\eta$ window based on the tracking system coverage of CMS and ATLAS~\cite{CMS_2006,CSC_2009}.
The efficiency of the \bjet s are lower than that of the WBF jets. 
The integrated jet reconstruction efficiency for each algorithm is given in Table \ref{T:jetreco-eff}.
\siscone7 exhibits inferior performance than the other three jet reconstruction algorithms whose performance are comparable.
\begin{table}[h]
\centering
\begin{tabular}{|c|c|c|c|c|}\hline \hline
Jet Algorithm & \siscone4 & \siscone7 & \antikt4 & \antikt6 \\ \hline
Four Jet Efficiency & 69.6\% & 58.2\% & 72.3\% & 70.1\%\\
\hline \hline
\end{tabular}
\caption{\label{T:jetreco-eff} Jet efficiency for tagging the four signal jets with their appropriate $\eta$ window.}
\end{table}
\\ \\
In the scenario in which two quarks are emitted with a small relative angle, we investigated the efficiency of identifying two jets as a function of their $\Delta R$ separation.
For this study we required all four outgoing quarks to have $\pt>15$ GeV/c, both b-quarks to be within $|\eta|<2.5$, and both WBF quarks to be within $|\eta|<4.5$. 
Figure~\ref{efficiency_dr_b} shows the efficiency to reconstruct two b quarks as a function of $\Delta R$. 
Figure~\ref{efficiency_dr_vbf} shows the efficiency to reconstruct two WBF quarks as a function of $\Delta R$. 
In both Figures~\ref{efficiency_dr_b} and~\ref{efficiency_dr_vbf}, the inefficiency at small $\Delta R$ is due to the cone radius/aggregation distance of the jet reconstruction algorithm. 
For the b-jets, the low efficiency at large $\Delta R$ is due to the correlation between $\Delta R$ and the low \pt\ \bjet s.\\ \\
Similar to the individual jet efficiency the \siscone7 configuration yields noticeably worse results than the other jet configurations.
This is particularly true in the case of the two \bjet\ pairing in which the large cone size is problematic given the reduced $\eta$ window.
Again, the other three jet configurations yield similar results.
\subsubsection{Linearity and Resolution}
\begin{figure}
\begin{center}
\includegraphics[width=0.45\textwidth]{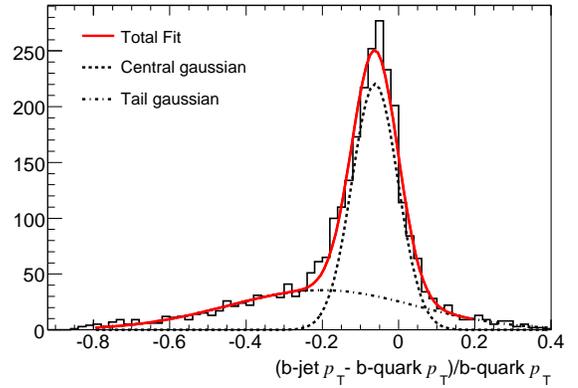}
\caption{Distribution of the \pt\ variable, defined in Equation~\ref{dET}, in one of the \pt\ slices of the b-quarks \mc\ sample. 
The result of the fit to the sum of two gaussians (Central gaussian + Tail gaussian) is also shown.}
\label{fig:fit}
\end{center}
\end{figure}

\begin{figure}
\begin{center}
\includegraphics[width=0.45\textwidth]{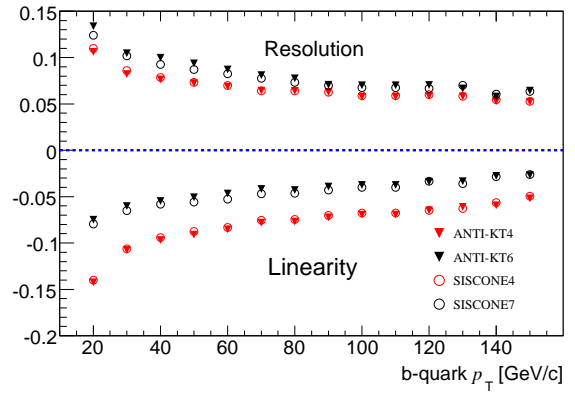}
\caption{\label{f:b_reso} Linearity and resolution (mean value and sigma of the central gaussian in the $\Delta \pt$ fits) as a function of the b-quark transverse energy. }
\end{center}
\end{figure}

\begin{figure}
\begin{center}
\includegraphics[width=0.45\textwidth]{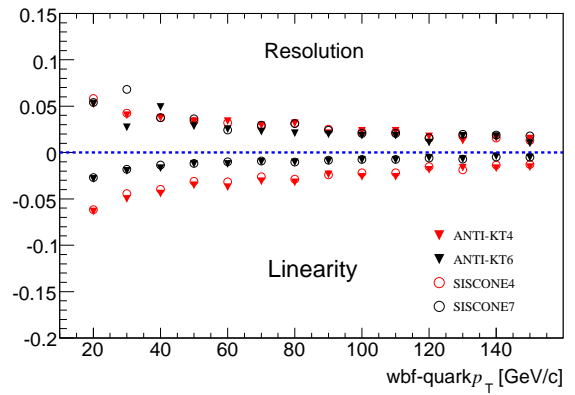}
\caption{\label{f:vbf_reso} Linearity and resolution (mean value and sigma of the central gaussian in the $\Delta \pt$ fits) as a function of the WBF transverse energy.}
\end{center}
\end{figure}

In addition to the efficiency, we considered the performance of the jet transverse momentum linearity and resolution for each jet configuration.
The jet transverse momentum linearity and  resolution was calculated using Equation~\ref{dET}.
\begin{equation}
\label{dET}
\Delta \pt= \frac{\pt^{j}-\pt^{p}}{\pt^{p}}
\end{equation}
Here $\pt^{j}$ and $\pt^{p}$ are the transverse momentum of the jet and outgoing parton, respectively.
We parameterized the distributions with a double Gaussian function and used the Gaussian with the largest amplitude to provide the linearity and the resolution. 
An example of the double Gaussian fit is shown in Figure \ref{fig:fit}. 
Figures \ref{f:b_reso} and \ref{f:vbf_reso} show the linearity and resolution as a function of the \pt\ of the b-quark and WBF-quarks, respectively. 
From these distributions, we conclude that the \siscone7 and the \antikt6 provide the best linearity results without any noticeable degradation of the resolution compared to the other two configurations. 
Additionally, we observed a significant performance degradation of the b-quarks jet reconstruction compared to WBF-quarks.
Consequently, additional corrections to the \bjet s are required and are summarized in Section~\ref{S:bjet_calib}.\\ \\
In conclusion,  the \siscone4, \antikt4, and \antikt6 configurations gave similar efficiency and resolution results, however, \antikt6 provides significant improvement in linearity.
Consequently, we used \antikt6 for the rest of this paper.
\subsection{Weak-Boson-Fusion Jet Identification}\label{S:VBF_ID}
\begin{figure}[hbt]
\begin{center}
\includegraphics[width=0.45\textwidth]{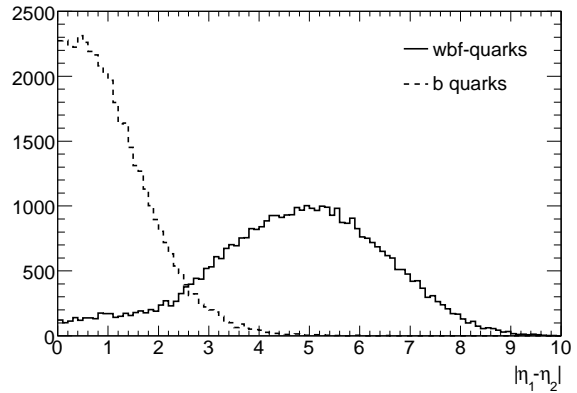}
\caption{Distribution of the absolute distance in pseudo-rapidity between WBF quark pair and b-quark pair for a 115~GeV/c$^{2}$ Higgs mass at 14 TeV \pp\ collisions.}
\label{F:lh_deta}
\end{center}
\end{figure}

\begin{figure}[hbt]
\begin{center}
\includegraphics[width =0.45\textwidth]{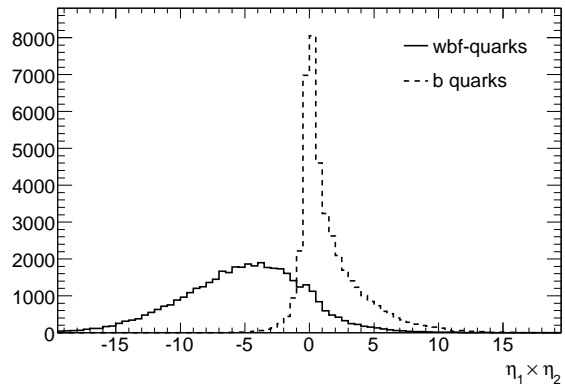}
\caption{Distribution of the product of pseudo-rapidities, $\eta_{1}\times\eta_{2}$, between WBF quark pair and b-quark pairfor a 115~GeV/c$^{2}$ Higgs mass at 14 TeV \pp\ collisions.}
\label{F:lh_peta}
\end{center}
\end{figure}

\begin{figure}[hbt]
\begin{center}
\includegraphics[width =0.45\textwidth]{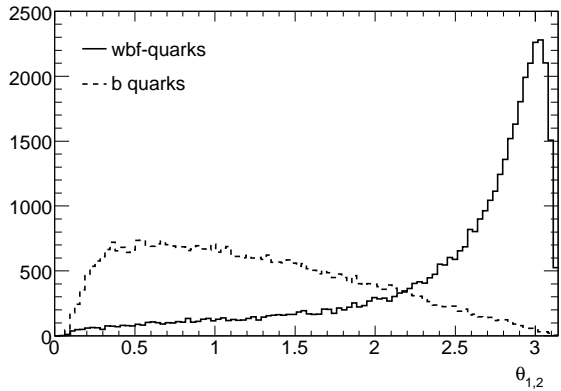}
\caption{Distribution of the 3-dimensional angle between the two jets ($\theta_{1,2}$) between WBF quark pair and b-quark pair for a 115~GeV/c$^{2}$ Higgs mass at 14 TeV \pp\ collisions.}
\label{F:lh_angle}
\end{center}
\end{figure}

\begin{figure}[hbt]
\begin{center}
\includegraphics[width =0.45\textwidth]{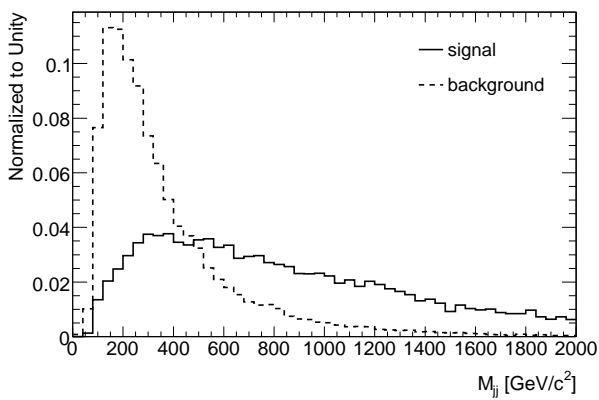}
\caption{Distribution of the invariant mass of the two WBF jets for a 115~GeV/c$^{2}$ Higgs mass at 14 TeV \pp\ collisions.}
\label{F:vbf_mjj}
\end{center}
\end{figure}

\begin{figure}[hbt]
\begin{center}
\includegraphics[width =0.45\textwidth]{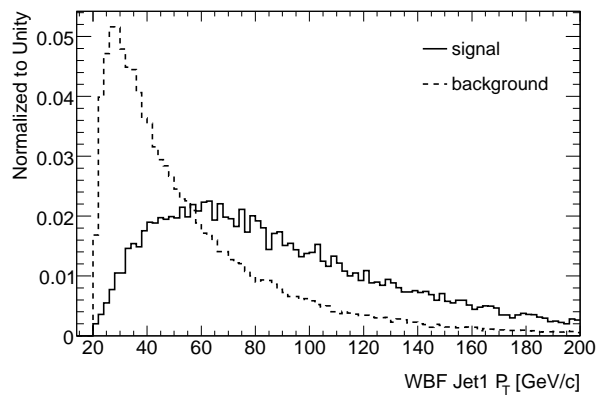}
\caption{Distribution of the transverse momentum of the highest transverse momentum WBF jet for a 115~GeV/c$^{2}$ Higgs mass at 14 TeV \pp\ collisions.}
\label{F:vbf_pt1}
\end{center}
\end{figure}

\begin{figure}[hbt]
\begin{center}
\includegraphics[width =0.45\textwidth]{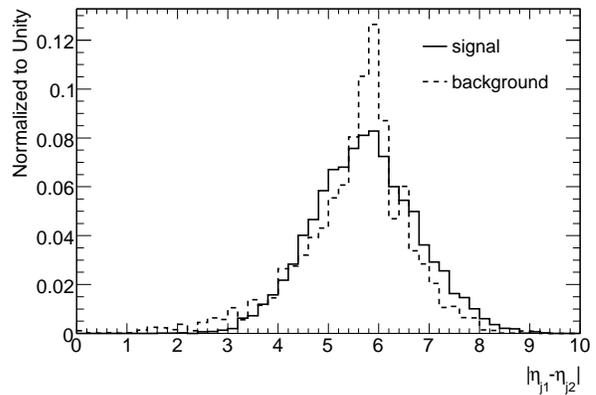}
\caption{Distribution of the absolute distance in pseudo-rapidity, $|\Delta \eta_{1,2}|=|\eta_{1}-\eta_{2}|$, between WBF jets for a 115~GeV/c$^{2}$ Higgs mass at 14 TeV \pp\ collisions.}
\label{F:vbf_deta}
\end{center}
\end{figure}

The geometrical topology of the WBF jets provides significant distinction between signal and background.
In addition, this topology provides a clean separation between the WBF jets from the \bjet s in signal events. 
This is particularly important when trying to reconstruct the Higgs mass from \bjet\ candidates since incorrectly identifying \bjet s degrades the Higgs mass resolution.\\ \\
Accordingly, we developed a WBF tagger which uses a likelihood ratio to distinguish the desired WBF jet pair from b-jet pairs within the signal.
The likelihood ratio, $y$, is defined using the WBF jet pair likelihood ($\mathcal{L}_{v}$) and the \bjet\ pair likelihood ($\mathcal{L}_{b}$), as shown in Equation \ref{E:lh_ratio}.
\begin{equation}\label{E:lh_ratio}
y= \frac{\mathcal{L}_{v}}{\mathcal{L}_{v}+\mathcal{L}_{b}}
\end{equation}
Here, $\mathcal{L}_{v/b}$, is defined as the product of 1-dimensional probability density functions, $P(x_{i})$, as shown in Equation \ref{E:lh}, where $x_{i}$ is the $i$th variable.
\begin{equation}\label{E:lh}
\mathcal{L}_{v/b}= \prod P_{v/b}(x_{i})
\end{equation}
Three geometrical variables were used in the likelihood ratio to identify the best WBF jet candidates.
These variables are listed below and are shown in Figures \ref{F:lh_deta}, \ref{F:lh_peta}, and \ref{F:lh_angle}.
\begin{enumerate}
\item The absolute difference in pseudo-rapidity, $|\Delta \eta_{ij}|=|\eta_{i}-\eta_{j}|$.
\item The product of pseudo-rapidities, $\eta_{i}\times\eta_{j}$.
\item The 3-dimensional angle between the two jets, $\theta_{ij}$.
\end{enumerate} 
Here $i$ and $j$ correspond to the pair of WBF and b quarks candidates.
The probability density functions used in the likelihood ratio were generated using matrix element distributions from an independent signal sample with a 115 GeV/c$^{2}$ Higgs mass.
The WBF tagger is indifferent to the jet flavour, as such, it occasionally selects \bjet s. 
We observed approximately $5\%$ \bjet\, contamination in the signal samples after all signal selection cuts
Once two WBF jet candidates are identified, the following kinematic cuts are applied.\\ 
\paragraph{Invariant Mass:}
The majority of the QCD multi-jet background will produce `soft' jets resulting in a low invariant mass compared to the two WBF jets from the signal, as shown in Figure \ref{F:vbf_mjj}.
Consequently, we required the invariant mass, $M_{jj}$, of the WBF jet candidates to be greater than 695 GeV/c$^{2}$ in order to reduce the QCD backgrounds.\\ \\
\paragraph{Transverse Momentum:}
We required the highest \pt\ WBF jet to have a \pt\ greater than 55 GeV/c to remove events with a `soft' tagged jet which are copiously produced in QCD background events.
The \pt\ distributions of  the highest \pt\ WBF jet for a 115~GeV/c$^{2}$ mass Higgs signal and the background are presented in Figure \ref{F:vbf_pt1}.\\ \\
\paragraph{Geometrical Distributions:}
Although the WBF tagger finds the best jet pair which satisfies the geometrical distributions of the signal WBF jets, the $\Delta \eta$ distribution, shown in Figure \ref{F:vbf_deta}, provides additional discrimination between signal and background.
Consequently, we required that the  $\Delta \eta$ of the WBF pair be greater than 3.25.
\subsection{B-jets Identification}\label{S:B_ID}
\begin{figure}[hbt]
\begin{center}
\includegraphics[width =0.45\textwidth]{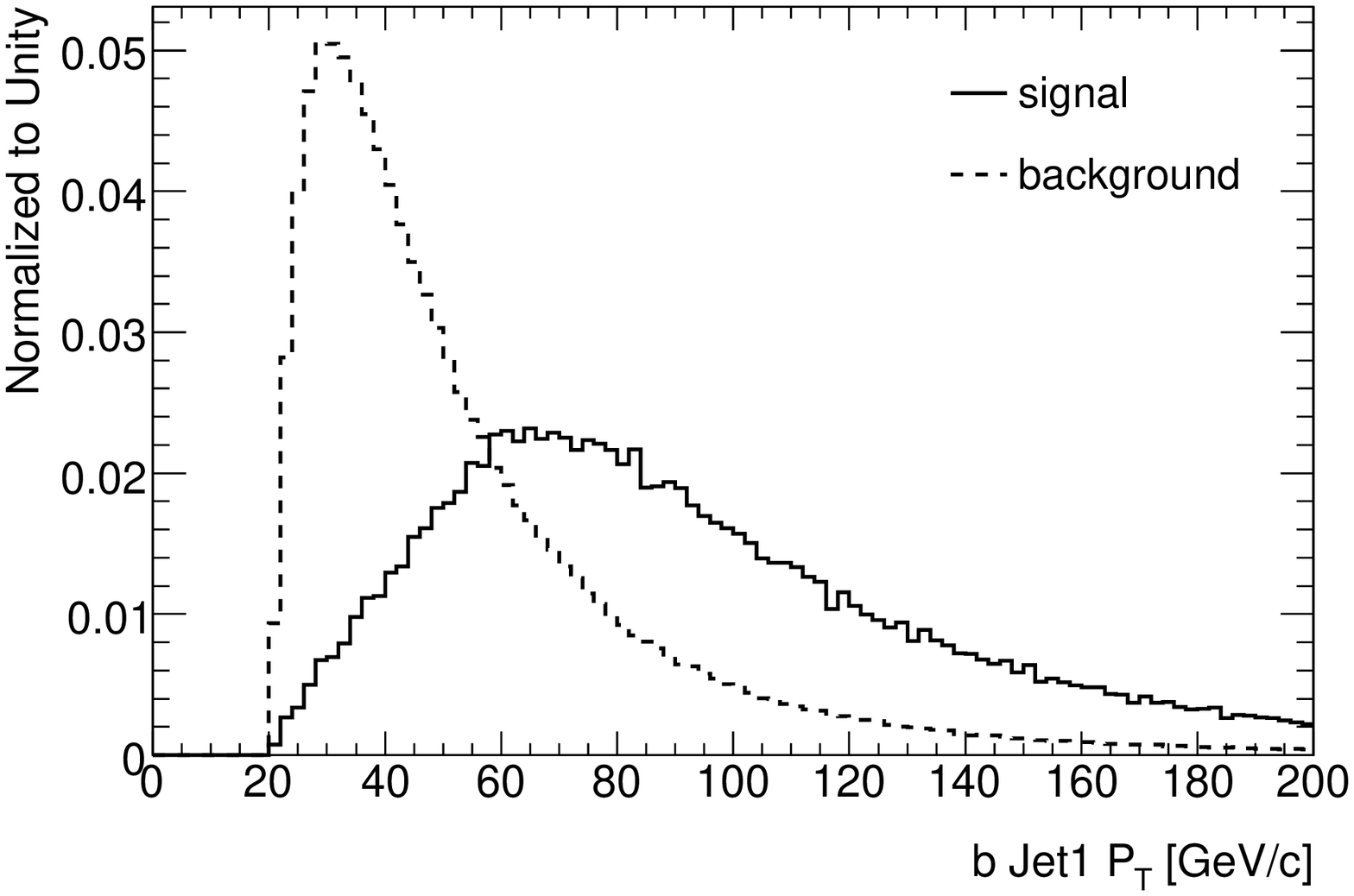}
\caption{Distribution of the transverse momentum of the highest transverse momentum \bjet\ for a 115~GeV/c$^{2}$ Higgs mass at 14 TeV \pp\ collisions.}
\label{F:b_pt1}
\end{center}
\end{figure}

\begin{figure}[hbt]
\begin{center}
\includegraphics[width =0.45\textwidth]{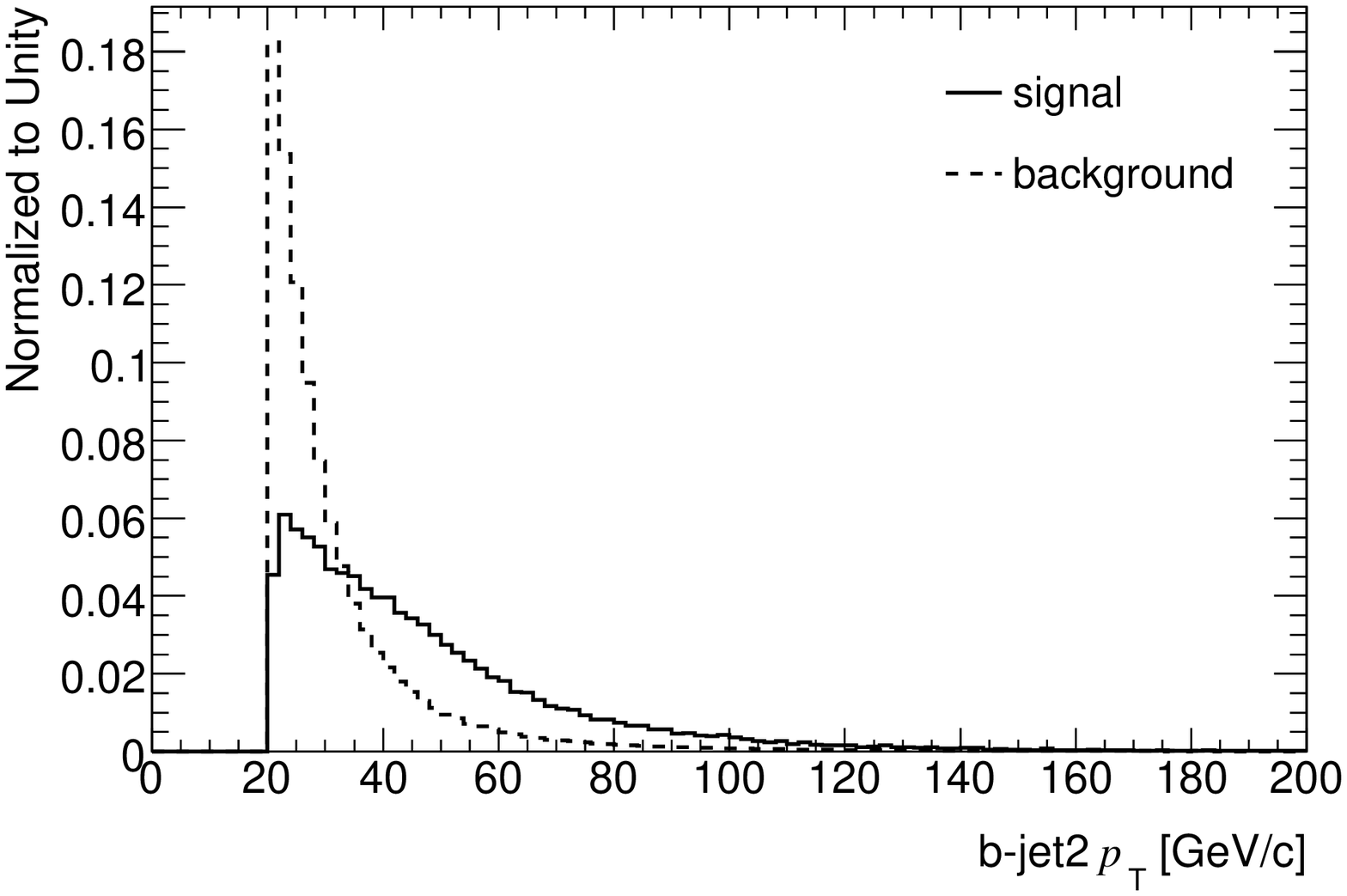}
\caption{Distribution of the transverse momentum of the second highest transverse momentum \bjet\  for a 115~GeV/c$^{2}$ Higgs mass at 14 TeV \pp\ collisions.}
\label{F:b_pt2}
\end{center}
\end{figure}

\begin{figure}[hbt]
\begin{center}
\includegraphics[width =0.45\textwidth]{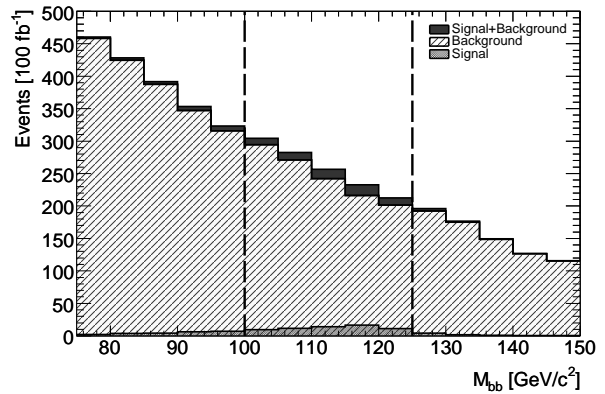}
\caption{Distribution of the invariant mass of the two \bjet s with all selection criteria itemized in Table~\ref{T:cutlist} with the additional requirement of \pt\ $>$ 20 GeV/c on both \bjet s  for a 115~GeV/c$^{2}$ Higgs mass at 14 TeV \pp\ collisions.}
\label{F:mbb_no_cut}
\end{center}
\end{figure}

\begin{figure}[hbt]
\begin{center}
\includegraphics[width =0.45\textwidth]{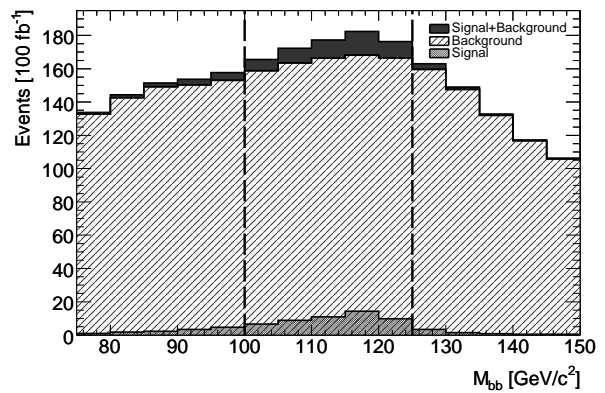}
\caption{Distribution of the invariant mass of the two \bjet s with all selection criteria itemized in Table~\ref{T:cutlist} after applying a \pt\ cut of 60 GeV/c on the highest \pt\ \bjet\ for a 115~GeV/c$^{2}$ Higgs mass at 14 TeV \pp\ collisions.}
\label{F:mbb_w_cut}
\end{center}
\end{figure}

\begin{figure}[hbt]
\begin{center}
\includegraphics[width =0.45\textwidth]{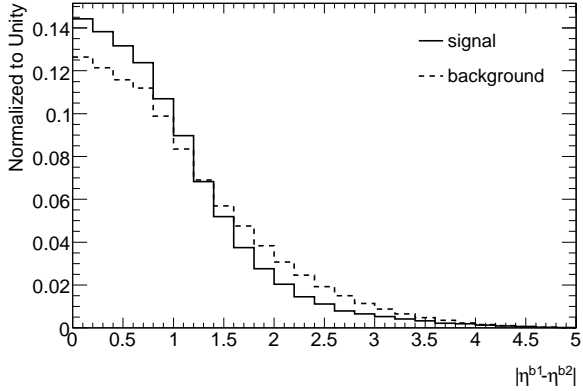}
\caption{Distribution of the absolute distance in pseudo-rapidity between \bjet s for a 115~GeV/c$^{2}$ Higgs mass at 14 TeV \pp\ collisions.}
\label{F:b_deta}
\end{center}
\end{figure}

\begin{figure}[hbt]
\begin{center}
\includegraphics[width =0.45\textwidth]{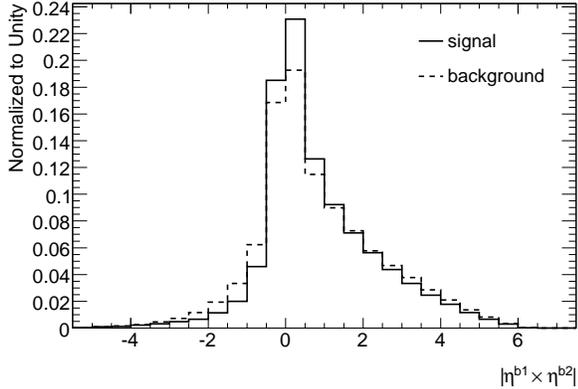}
\caption{Distribution of the product of pseudo-rapidity between \bjet s for a 115~GeV/c$^{2}$ Higgs mass at 14 TeV \pp\ collisions.}
\label{F:b_peta}
\end{center}
\end{figure}

\begin{figure}[hbt]
\begin{center}
\includegraphics[width =0.45\textwidth]{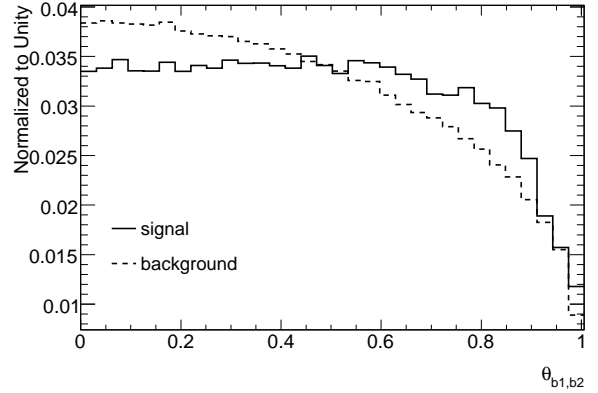}
\caption{Distribution of the cosine of the angle between the boost and decay axis of the $H\to b\bar b$ candidates for a 115~GeV/c$^{2}$ Higgs mass at 14 TeV \pp\ collisions.}
\label{F:b_angle}
\end{center}
\end{figure}

In order to estimate the expected b-jet identification at the LHC experiments, we assumed 60$\%$ b-tagging efficiency within $|\eta|<2.5$.
This b-tagging efficiency corresponds to approximately a 10 and 200 rejection factor for \ljet s\footnote{Jets originating from u,d, and s quarks and gluons.} and \cjet s, respectively~\cite{CSC_2009}.\\ \\
We simulated the expected performance of the b-tagging by first matching jets which originate from the \me\,b-quark.
Once the jets are matched, we randomly generated the efficiency using a uniform distribution.
Jets which passed based on the assumed b-tagging efficiency were identified as a \bjet.
Similarly, we simulated the expected fake rate from light and c jets by using the same prescription assuming the aforementioned rejection factors.
In order to retain the full \mc\ statistics, we required that each event produce at least two \bjet s.
Consequently, we kept count of the number of iterations required to satisfy this criteria and we normalized the expected number of events accordingly.
Once the \bjet s were identified, we applied the cuts listed below.\\
\paragraph{Transverse Momentum:}
Similar to the WBF jets, the QCD background has relatively `soft' \pt\ \bjet s compared to those in the signal events, as shown in Figures \ref{F:b_pt1} and  \ref{F:b_pt2}.
However, applying a high \pt\ cut on the \bjet\ candidates significantly shapes the invariant mass of the two \bjet s in the background, as shown in Figures \ref{F:mbb_no_cut} and \ref{F:mbb_w_cut}.
Consequently, we only applied a low  \pt\ cut of 20 GeV/c on both \bjet\ candidates.\\ \\
\paragraph{Geometrical Distributions:}
In contrast to the WBF jets, the \bjet s are generally close in $\eta$ as shown in Figure \ref{F:b_deta}.
Accordingly, we required that the \bjet s are within $\Delta \eta <1.25$.
Additionally, we apply a cuts on the $\eta$ product of the two \bjet and the angle between the boost and the decay axis of the $H\to b \bar{b}$ which are shown in Figure \ref{F:b_peta} and \ref{F:b_angle} respectively.\\ \\
\paragraph{Invariant Mass:}
The most discriminating variable in this analysis is the invariant mass of the two \bjet s.
Signal events peak at the nominal Higgs mass while the QCD background has a broad continuum distribution as shown in Figure \ref{F:mbb_no_cut}.
In order to estimate the significance of this analysis, we counted candidates in a two \bjet\ invariant mass window near the nominal signal Higgs mass. The lower limit of the mass window is 100, 108, and 117~GeV/$c^2$ and the upper limit of the mass window is 125, 136, and 147~GeV/$c^2$ optimized for a Higgs mass of 115, 125 and 135~GeV/$c^2$, respectively.

\subsection{B-jet Calibration}\label{S:bjet_calib}
\begin{figure}
\begin{center}
\includegraphics[totalheight=0.3\textwidth]{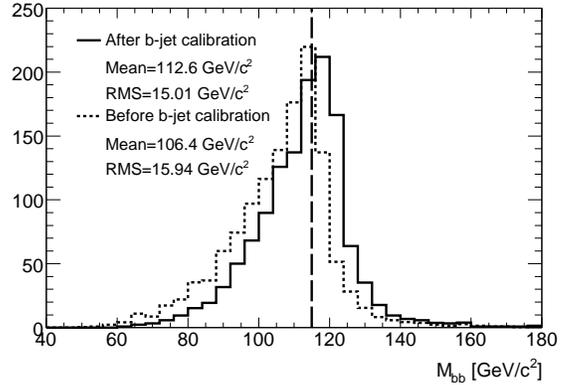}
\caption{Reconstruction of the invariant mass of the $b\bar{b}$ pairs for a 115~GeV/c$^{2}$ Higgs mass at 14~TeV \pp\ collisions.}
\label{bcac}
\end{center}
\end{figure}

The results of this analysis strongly rely on the ability to identify jets originating from b-quarks and accurately reconstruct their 4-momentum.
However, the \bjet\ linearity and resolution are noticeably worse than light-quark jets as described in Section \ref{S:JET_PERF}.
Here, we present a calibration technique to improve the \bjet\ performance which consequently corrects the invariant mass of the two \bjet s.\\ \\
We have implemented the {\em numerical inversion} method which corrects the transverse momentum of the \bjet s by applying a multiplicative scale factor derived from the ratio of the transverse momentum between the reconstructed jet and the original b-quark.
The scale factors were derived in bins of $\eta$ and \pt\ using the \bbjph\ and \bbjjph\ background samples.
Additionally, two sets of calibration scale factors were derived for \bjet s with and without matching muons. A muon is considered matched to a jet if $\Delta R_{j\mu}<0.4$ with $p_T(\mu)$ greater than 5 GeV/c.
The result of the calibration gave mean values of the invariant mass of the signal events with a Higgs mass of 115 GeV/c$^2$
shifts from 106.4 GeV/c$^2$ to 112.6 GeV/c$^2$ while slightly reducing the relative RMS from 14.98 \% to 13.33 \%.
We have applied this method to our b-jets to improve the result on the reconstruction of invariant mass resonances. 
In Figure~\ref{bcac}, we can see the difference in a 115 GeV/c$^2$ Higgs mass reconstruction before and after the calibration.
\subsection{Central Jet Veto}\label{S:CJV}
A key aspects of the WBF Higgs process is its electroweak nature.
At leading-order the signal process does not have any color flow between the interacting quarks.
In contrast, the majority of the backgrounds have QCD radiation between interacting quarks.
Removing events with jet activity in the region between the WBF jets provides a potentially powerful tool to suppress backgrounds.
In addition to the hard interaction of interest, each event includes an underlying event,
consequently, the choice of showering, hadronization, and UE models employed in the Monte Carlo samples can potentially impact the efficacy of the central jet veto (CJV).
We investigated two techniques to optimize the central jet veto.
\begin{itemize}
\item {\it Fixed $\eta$ Window:} Events are vetoed if additional jets above a given \pt\ threshold are found within a fixed $\eta$ window.
\item {\it Dynamic $\eta$ Window:} Similar to the fixed $\eta$ window, however, the $\eta$ window is defined by the $\eta$ of the two WBF jets event-by-event.
\end{itemize}
The results from this study were determined using a 115 GeV/c$^{2}$ Higgs mass signal sample and the \bbjjph\ background samples.
Both samples were produced using the nominal {\it Perugia} tune.
For the {\it fixed} $\eta$ window and the , we scanned the $\pt \times |\eta|$ plane and found that a veto on jets within $|\eta|<2$ and \pt\ greater than 25 GeV/$c$ was optimal.
For the {\it dynamic} $\eta$ window, we scanned the \pt\ threshold and found the optimal value to be 25 GeV/c. The {\it fixed} $\eta$ window technique provides approximately a 16\% increase in signal significance - a modest  5\% improvement over the {\it dynamic} $\eta$ window technique.


\section{Systematic Uncertainties}\label{S:sys_uncertainty}
In this section, we considered several sources of systematic uncertainties associated with the signal and background cross-sections and the choice of \mc\ tunes.
\subsection{Factorization/Renormalization Scale}
For the analysis, the nominal parametrization for the renormalization/factorization scale is listed in Section~\ref{S:Evg}.
Within the \alpgen\ program, the renormalization/factorization scales are linearly correlated ($\mu_{\rm{F}}=\mu_{\rm{R}}= \mu$).
As such, anti-correlated relationship between \fr\, scales was not considered.
In order to estimate the uncertainty related to the parametrization of the scales, we scanned the various \alpgen\ parametrizations and multiplied the parameterization by 0.5 and 2.
The result from these variations for a 115 GeV/c$^{2}$ Higgs mass were within $5\%$ of the results using the nominal settings.
Additionally, we performed the same methodology on the background and the results are shown in Table~\ref{t:vbfph_theory_14TeV}.
\subsection{Parton Density Function}
The parton distribution functions (PDFÕs) of the nucleon are central to determining the cross-section of the processes at proton-proton collisions. 
We used the full group of 40 CTEQ6M sets which allowed the determination of the PDF systematics uncertainty using the prescription defined in~\cite{PDF_2007} using the Equation \ref{E:PDF_Master_a}.
\begin{equation}\label{E:PDF_Master_a}
\Delta \sigma = \frac{1}{2} \sqrt{ \sum_{i=1}^{20} (\sigma_{2i}-\sigma_{2i-1})^{2} },\\
\end{equation}
Here $\sigma_{i}$ are the cross-sections based on the $i^{th}$ PDF sets. 
For the Higgs process we found that the contribution for the PDF uncertainty is of the order of $\sim 4\%$. \\ \\
From the different background processes, the PDF uncertainty ranges from 3.52\% to 3.89\% and are shown in Table \ref{t:vbfph_theory_14TeV}. \\ \\
\begin{table}[htdp]
\begin{center}
\begin{tabular}{|c|c|c|c|}\hline \hline
\multirow{2}{*}{Number of Partons}  &  \multirow{2}{*}{$\sigma$ [pb]} &  \multirow{2}{*}{$\frac{\Delta \sigma_{scale}}{\sigma}$ [$\%$]} & \multirow{2}{*}{$\frac{\Delta \sigma_{PDF}}{\sigma}$ [$\%$]}\\ 
 &&& \\  \hline
\multicolumn{4}{|c|}{\multirow{2}{*}{$b\bar{b}$+$n$parton+$\gamma$}} \\
\multicolumn{4}{|c|}{}  \\  \hline
\multirow{2}{*}{1} & \multirow{2}{*}{1088}  & +34.5 & \multirow{2}{*}{3.89}\\
& & -38.8  & \\
\multirow{2}{*}{2} & \multirow{2}{*}{658}  & +52.4 & \multirow{2}{*}{3.58}\\
&  & -50.8 &\\
\hline
\multicolumn{4}{|c|}{\multirow{2}{*}{$n$parton+$\gamma$}} \\
\multicolumn{4}{|c|}{}  \\  \hline
\multirow{2}{*}{3} & \multirow{2}{*}{45789} & + 36.1 & \multirow{2}{*}{3.78} \\
& &  --22.8 & \\
\multirow{2}{*}{4} & \multirow{2}{*}{17595} & + 59.7 & \multirow{2}{*}{3.16} \\
& & --31.2 & \\
\hline
\multicolumn{4}{|c|}{\multirow{2}{*}{$Z(\to q \bar{q})$+$n$parton+$\gamma$}} \\
\multicolumn{4}{|c|}{}  \\  \hline
\multirow{2}{*}{1} & \multirow{2}{*}{27} & +15.9 & \multirow{2}{*}{3.52}\\
&  & -12.1 & \\
\multirow{2}{*}{2} & \multirow{2}{*}{18} & + 32.5 & \multirow{2}{*}{3.73}\\
& & -24.4 & \\
\hline \hline
\end{tabular}
\caption{\label{t:vbfph_theory_14TeV} Nominal cross section, factorization and renormalization scales, and PDF uncertainties for backgrounds \mc\, at 14 TeV \pp\ collisions.}
\end{center}
\end{table}
\subsection{Showering, Hadronization, and the Underlying Event Model}\label{S:UE_SYST}
\begin{figure}
\begin{center}
\includegraphics[width=0.48\textwidth]{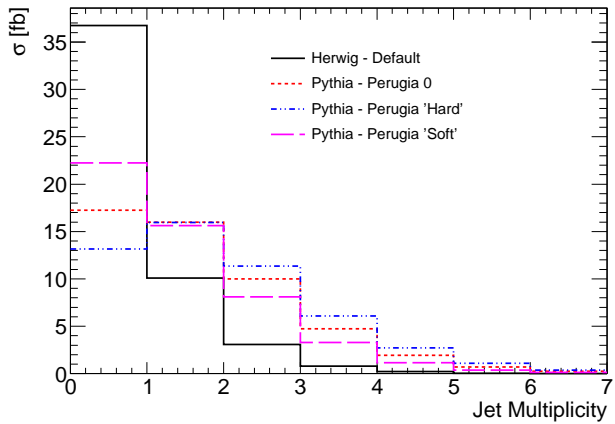}
\caption{\label{f:signal_nextajets} Comparison of the jet multiplicity for jet not originating from the hard process between \herwig\ and the \pythia\ tunes for a 115~GeV/c$^{2}$ Higgs mass at 14 TeV \pp\ collisions.}
\label{efficiency}
\end{center}
\end{figure}

\begin{figure}
\begin{center}
\includegraphics[width=0.48\textwidth]{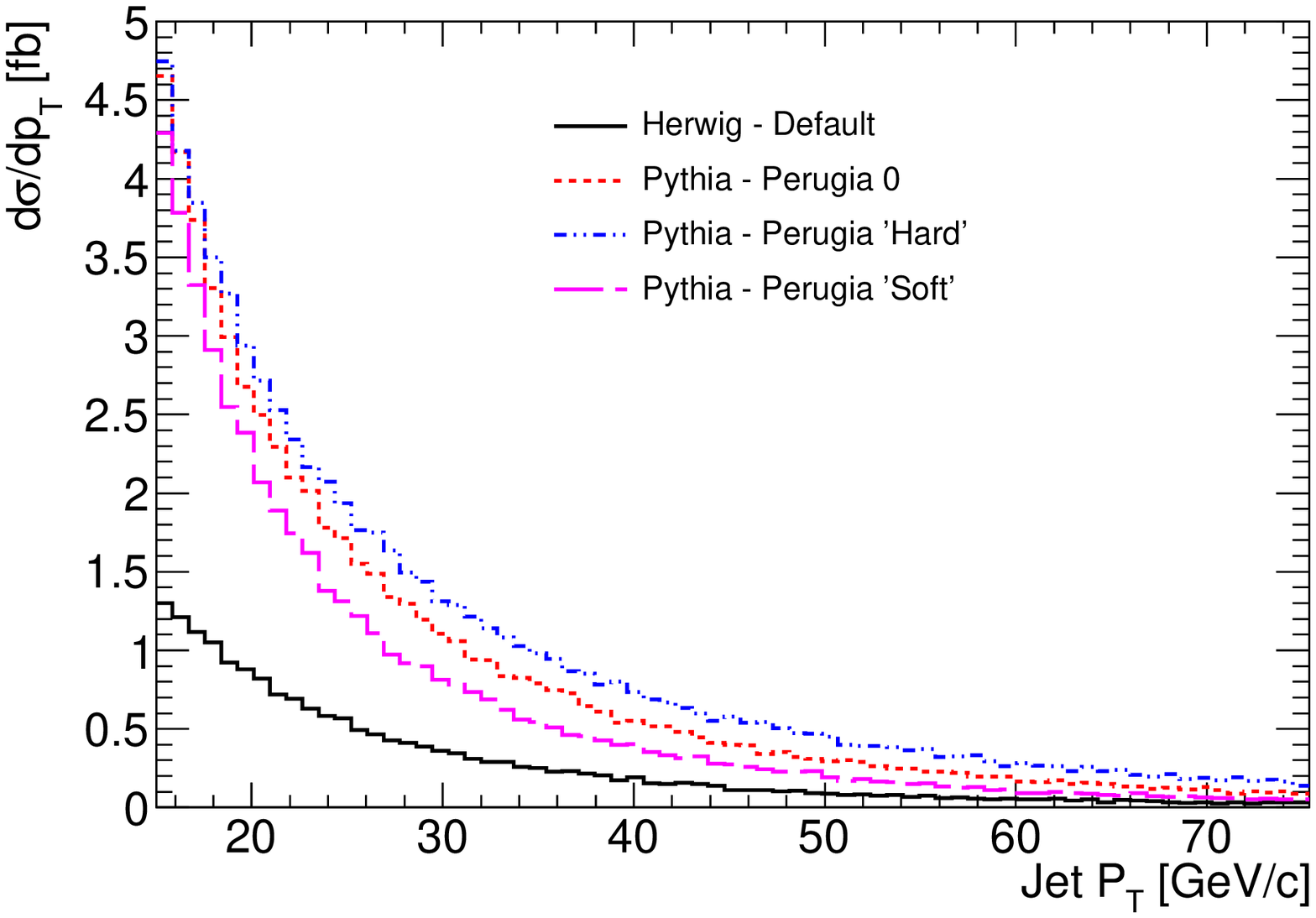}
\caption{\label{f:signal_jetspt} Comparison of the jet \pt\ distribution for jet not originating from the hard process between \herwig\ and the \pythia\ tunes for a 115~GeV/c$^{2}$ Higgs mass at 14 TeV \pp\ collisions.}
\label{efficiency}
\end{center}
\end{figure}

\begin{figure}
\begin{center}
\includegraphics[width=0.48\textwidth]{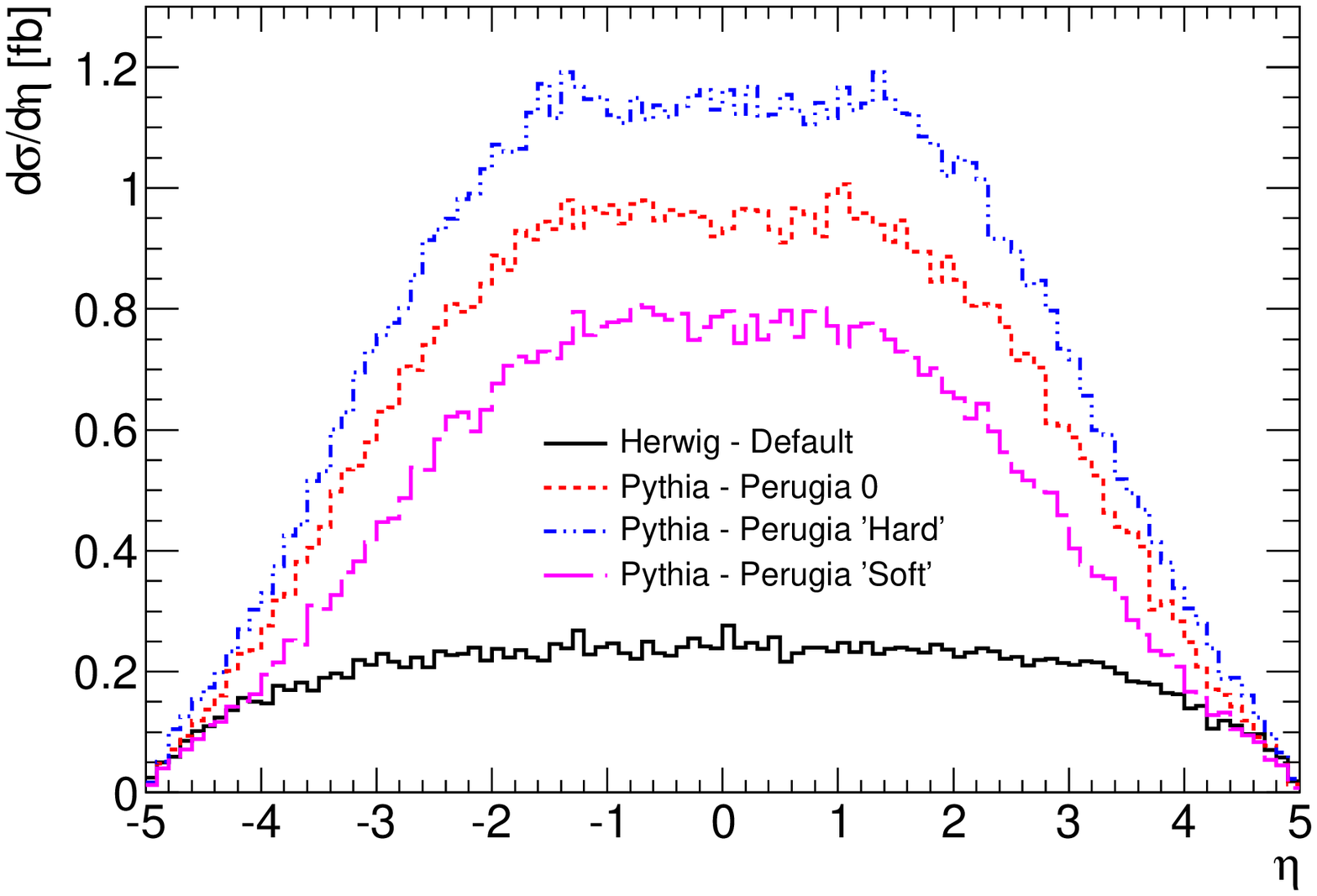}
\caption{\label{f:signal_jetseta} Comparison of the jet $\eta$ distribution for jet not originating from the hard process between \herwig\ and the \pythia\ tunes  for a 115~GeV/c$^{2}$ Higgs mass at 14 TeV \pp\ collisions.}
\label{efficiency}
\end{center}
\end{figure}

\begin{figure}
\begin{center}
\includegraphics[width=0.48\textwidth]{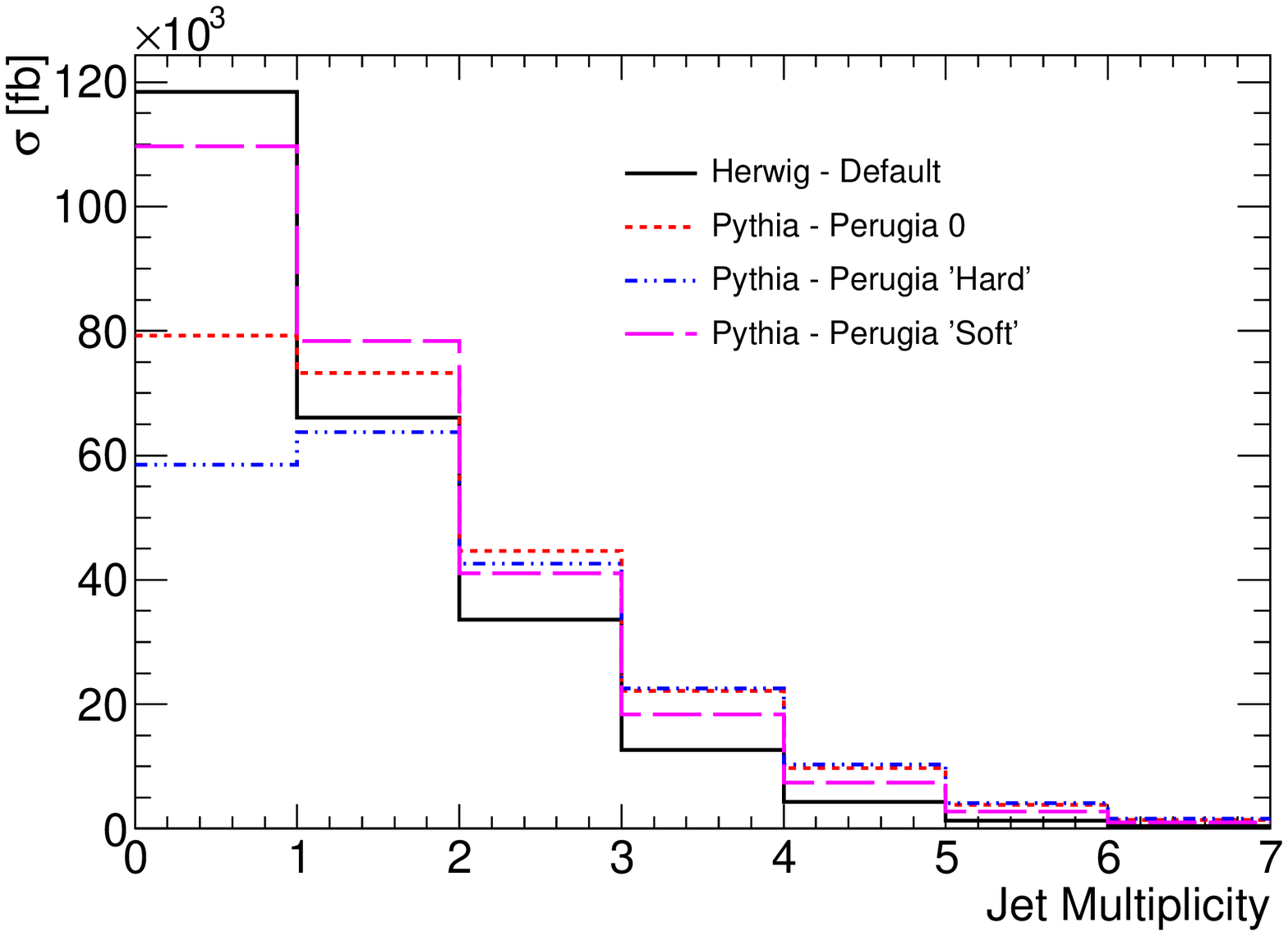}
\caption{\label{f:bkg_nextajets} Comparison of the jet multiplicity for jet not originating from the hard process between \herwig\ and the \pythia\ tunes for \bbjjph\ events at 14 TeV \pp\ collisions.}
\label{efficiency}
\end{center}
\end{figure}

\begin{figure}
\begin{center}
\includegraphics[width=0.48\textwidth]{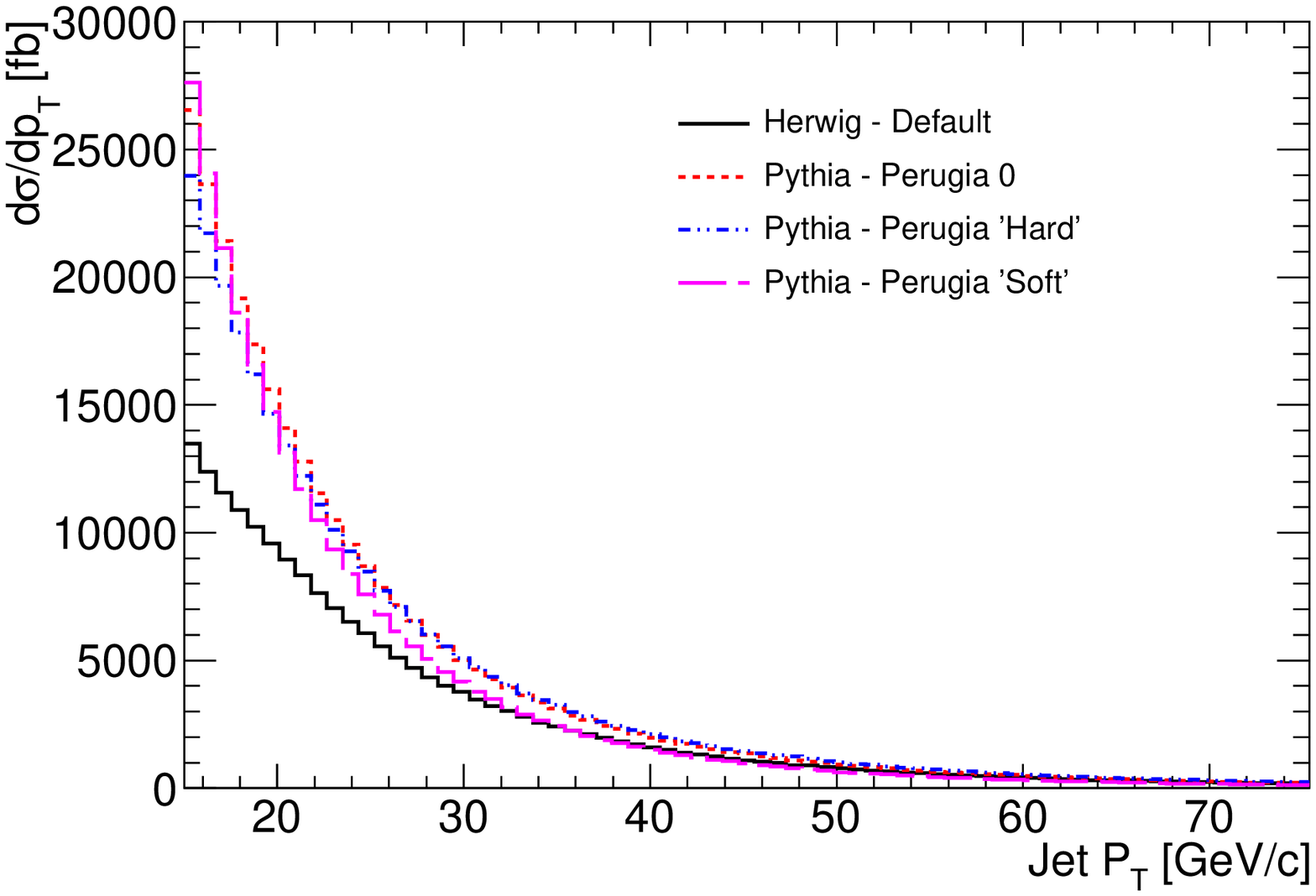}
\caption{\label{f:bkg_jetspt} Comparison of the jet \pt\ distribution for jet not originating from the hard process between \herwig\ and the \pythia\ tunes for \bbjjph\ events at 14 TeV \pp\ collisions.}
\label{efficiency}
\end{center}
\end{figure}

\begin{figure}
\begin{center}
\includegraphics[width=0.48\textwidth]{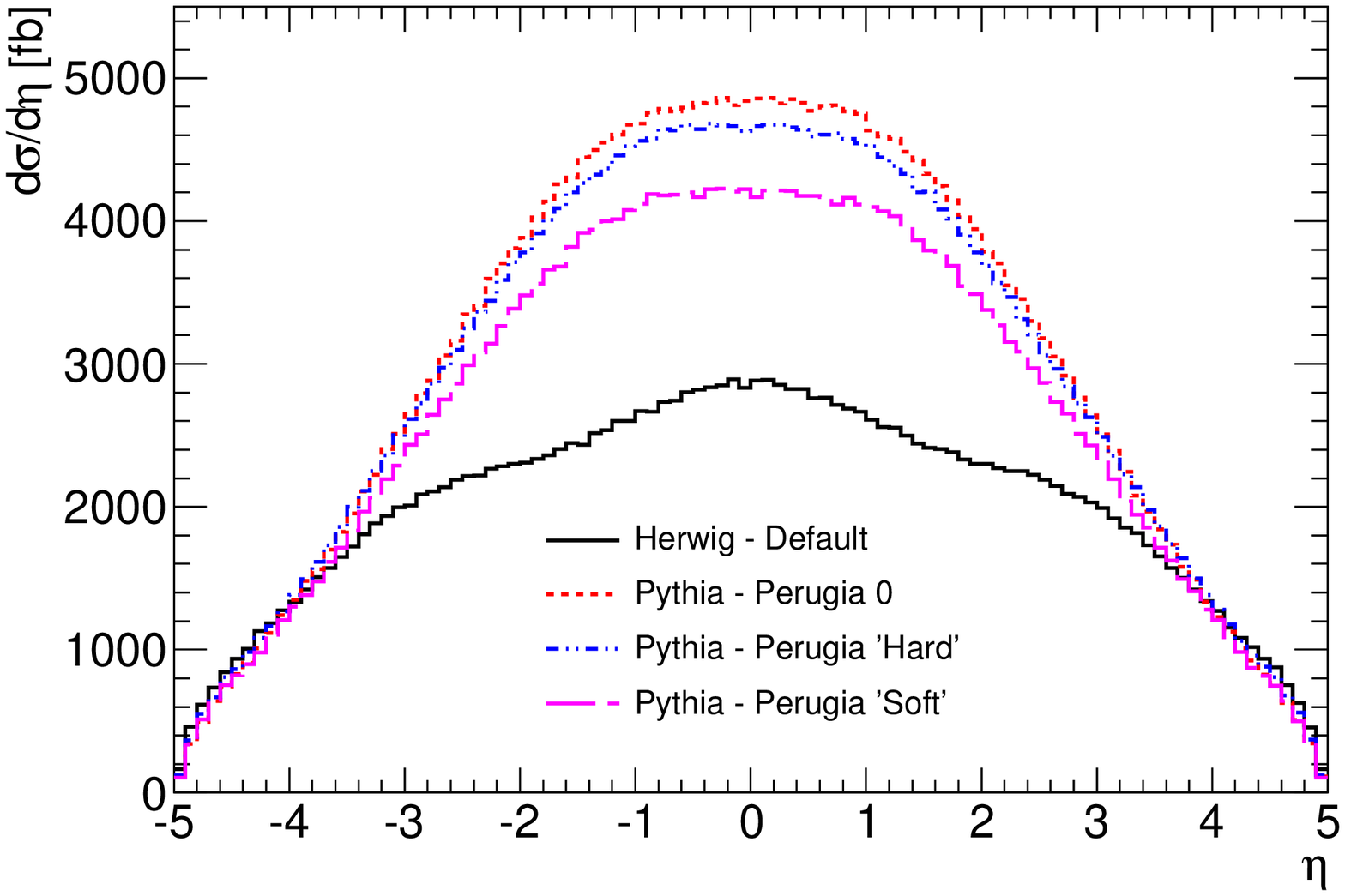}
\caption{\label{f:bkg_jetseta} Comparison of the jet $\eta$ distribution for jet not originating from the hard process between \herwig\ and the \pythia\ tunes for \bbjjph\ events  at 14 TeV \pp\ collisions.}
\label{efficiency}
\end{center}
\end{figure}

In order to estimate the systematic uncertainty originating from showering, hadronization, and the UE model, we processed the signal and background \alpgen\, 4-vectors using three variations of the {\it Perugia} Tunes which have been available since \pythia\, 6.4.20~\cite{Pythia64_2006}.
We considered the {\it Perugia} 0 as the nominal tune for this analysis and the {\it Perugia} `soft' and `hard' tunes for this systematic study.
The  {\it Perugia} variations are said to provide ``uncertainty bands'' and indicate an uncertainty of approximately 15$\%$ or less on the nominal tune~\cite{PerugiaTunes_09}.\\ \\
As stated in Section \ref{S:Evg}, the MLM prescription provides the ability to merge samples from different hard-parton multiplicity without double counting.
However, the MLM prescription is sensitive to the \mc\ tune.
Consequently, the {\it Perugia} tunes provide an estimation of the uncertainty of the MLM efficiency, as shown in Table \ref{t:vbfph_tune_14TeV}.
From the different background processes, the MLM efficiency uncertainty ranges from 10$\%$ to 20$\%$.\\ \\
Additionally, we investigated the properties of spurious jets between \herwig\ and the \pythia\ tunes within the signal and \bbjjph\ samples.
Comparing the multiplicity for spurious jets between \herwig\ and the \pythia\ tunes a noticeable difference we observed, as shown in Figures \ref{f:signal_nextajets} and  \ref{f:bkg_nextajets}.
For the \herwig\ tune, approximately 70\% of signal events and 50\% of  \bbjjph\ events do not have any additional jets while for the \pythia\ tunes both signal and background have approximately 40\% of events in the zero jet bin.
From Figures \ref{f:signal_jetspt} and \ref{f:bkg_jetspt}, we can see that the \pt\ distribution in signal and background are different between \herwig\ and \pythia\ which provides a more effective CJV using \herwig.
Additionally, there are 3.5-6.0 times more spurious central jets in the \pythia\ signal sample compared to \herwig, as shown in Figure \ref{f:signal_jetseta}.
In contrast, in the \bbjjph\ sample there is only 1.5-2.0 times more spurious central jets in \pythia\ compared to \herwig, as shown in Figure \ref{f:bkg_jetseta}.
Consequently, the results of our study indicate that the original estimation~\cite{VBFGamma07} on the efficacy of the CJV was to optimistic and only a modest improvement in significance is expected.
\\ \\
In summary, the {\it Perugia} tunes affect the expected number of signal and background events, as shown in Table \ref{t:vbfph_sig_tune_14TeV}, yielding an uncertainty of 15.6\% and 25.5\%, respectively.
However, the overall effect on the signal significance is approximately 4\%, as shown in Table \ref{t:vbfph_sig_tune_14TeV}.
\\
\begin{table}[htd]
\begin{center}
\begin{tabular}{|c|c|c|c|c|}\hline \hline
Number of  & \multicolumn{3}{c|}{MLM Efficiency [$\%$]}& \multirow{2}{*}{$\frac{\Delta \sigma_{MLM}}{\sigma}$ [$\%$]}\\ 
Partons       & Nominal & Soft & Hard & \\  \hline
\multicolumn{5}{|c|}{\multirow{2}{*}{$b\bar{b}$+$n$parton+$\gamma$}} \\
\multicolumn{5}{|c|}{}  \\  \hline
1 & 37.9 & 29.5 & 44.6 & 20.2\\
2 & 35.7 & 31.0 & 39.4 & 11.8\\
\hline
\multicolumn{5}{|c|}{\multirow{2}{*}{$n$parton+$\gamma$}} \\
\multicolumn{5}{|c|}{}  \\  \hline
3 & 20.5 & 17.2 & 25.3 & 19.3\\
4 & 17.8 & 15.3 & 21.0 & 15.8\\
\hline
\multicolumn{5}{|c|}{\multirow{2}{*}{$Z(\to q \bar{q})$+$n$parton+$\gamma$}} \\
\multicolumn{5}{|c|}{}  \\  \hline
1 & 40.6 & 34.2 & 47.0 & 15.8\\
2 & 44.5 & 39.8 & 48.6 & 10.0\\
\hline \hline
\end{tabular}
\caption{\label{t:vbfph_tune_14TeV} Background MLM uncertainties originating from \mc\ tunes for 14 TeV \pp\ collisions.}
\end{center}
\end{table}
\begin{table}[h]
\begin{center}
\begin{tabular}{|c|c|c|c|}\hline \hline
Estimated & \multicolumn{3}{c|}{{\it Perugia} Tunes}\\ 
Results & Nominal 	& Soft 		& Hard \\  \hline
Signal Events ($S$)		& 68$\pm$1 	& 56$\pm$1 	& 78$\pm$ 2 \\
Background Events ($B$)	 	& 1336$\pm$35 & 1072$\pm$33 	& 2013$\pm$53 \\
$S/\sqrt{B}$	 	& 1.86$\pm$0.06 & 1.71$\pm$0.06 & 1.74$\pm$0.05 \\
\hline \hline
\end{tabular}
\caption{\label{t:vbfph_sig_tune_14TeV} Expected signal and background events and overall significance uncertainties originating from \mc\ tunes for a 115 GeV/c$^{2}$ Higgs mass at 14 TeV \pp\ collisions. Only \mc\ statistical errors are quoted.}
\end{center}
\end{table}
\section{Results}\label{S:results}
Tables \ref{T:mh115_results}, \ref{T:mh125_results}, and \ref{T:mh135_results} give the expected number of events for 100 fb$^{-1}$ of data at 14 TeV collision energy after the application of successive cuts in the event selection outlined in Section \ref{S:ana} and listed in Table~\ref{T:cutlist} for a 115 GeV/c$^{2}$, 125 GeV/c$^{2}$, 135  GeV/c$^{2}$ Higgs mass, respectively.
Additionally, the signal significance derived from the aforementioned tables are shown in Table \ref{t:vbfph_sig_mass}. 
\begin{table}[htdp] 
\begin{center} 
\begin{tabular}{|c|c|}\hline \hline
Cut \# & Selection Criteria \\ \hline
1 & \pt($\gamma$) $> 30$ GeV/c \\ 
2 & \# of Jets $\geq 4$ \\
3 & \# of Central Jets $\geq 2$ \\
4 & Two b-jets  \\
5 & \pt$(j1)>$55 GeV/c \\ 
6 & M$(j1,j2)>$695.0 GeV/c$^2$ \\
7 & $\Delta \eta(j1,j2)>3.25$ \\
8 & $\theta(b1,b2)<0.92$\\
9 & $\Delta \eta(b1,b2)<1.25$\\
10 & $\eta(b1) \times \eta(b2)>-0.25$\\ 
11 &  \\
~~~i) $m_{h}=115$ GeV/c$^{2}$ &  $m_{bb}>100$ GeV/c$^{2}$\\
~~~ii) $m_{h}=125$ GeV/c$^{2}$ &  $m_{bb}>108$ GeV/c$^{2}$ \\
~~~iii) $m_{h}=135$ GeV/c$^{2}$ &  $m_{bb}>117$ GeV/c$^{2}$ \\
12 & \\
~~~i) $m_{h}=115$ GeV/c$^{2}$ &  $m_{bb}<125$ GeV/c$^{2}$\\
~~~ii) $m_{h}=125$ GeV/c$^{2}$ &  $m_{bb}<136$ GeV/c$^{2}$ \\
~~~iii) $m_{h}=135$ GeV/c$^{2}$ &  $m_{bb}<147$ GeV/c$^{2}$ \\
13 & Central Jet Veto   \\ \hline \hline
\end{tabular}
\caption{\label{T:cutlist} Selection Criteria. Here cuts 11 and 12 are apply to the appropriate Higgs mass sample.}
\end{center}
\end{table}
\begin{table}[h]
\begin{center}
\begin{tabular}{|c|c|c|c|}\hline \hline
Estimated & \multicolumn{3}{c|}{Higgs Mass}\\ 
Results & 115 	& 125 & 135 \\  \hline
Signal Events ($S$)		& 68$\pm$1 	& 58$\pm$1 	& 33$\pm$1 \\
Background Events ($B$)	 	& 1337$\pm$35 & 1341$\pm$38 & 1210$\pm$37 \\
$S/\sqrt{B}$	 	& 1.86$\pm$0.06 & 1.58$\pm$0.05 & 0.95$\pm$0.04 \\
\hline \hline
\end{tabular}
\caption{\label{t:vbfph_sig_mass} Expected signal and background events and overall significance uncertainties for Higgs mass of 115, 125, and 135~GeV/c$^{2}$ for 14 TeV \pp\ collisions. Only \mc\ statistical errors are quoted.}
\end{center}
\end{table}
\section{Summary}\label{S:summary}
In this paper, we presented the prospect for observing a light SM Higgs boson decaying to two b-quarks via WBF production with an associated photon at the LHC for 14 TeV collision energies.
We analyzed the signal and primary backgrounds after showering, hadronization, and included a UE model.
We studied various jet algorithms and concluded that \antikt6 provided the best overall performance.
We developed a likelihood WBF tagger in order to distinguish the Higgs \bjet s and the WBF jets.
We simulated the \bjet\ efficiency and light jet fake rate based on the latest expected detector performance studies.
For the \bjet\ candidates, we performed a calibration using the {\it numerical inversion} technique in order to provide a reconstructed invariant mass at the nominal Higgs mass.  \\ \\
Three sources of uncertainty were investigated: choice of \fr, \mc\ tunes, and input PDFs.
The largest uncertainty in this analysis was the background cross-section of the \bbjjph\ background which varies by approximately $\pm 50 \%$.
Consequently, for a 115 GeV/c$^{2}$ mass Higgs the significance can potentially be as large as $\sim$2.6 assuming a more conservative \fr\ scale.
The choice of \mc\ and \mc\ tune has several effects on the analysis such as the overall number of expected events in signal and background and efficacy of CJV.
We found that the previous estimation on efficacy of the CJV using \herwig\ was too optimistic.
The CJV using \herwig\  provides a 49\% increase in the signal significance while \pythia\ only increase the signal significance by 16\%.
Moreover, based on the latest ATLAS charge multiplicity results \cite{ATLASTunes_10} we conclude that the proper \mc\ tune is between the {\it Perugia} nominal and ``hard'' tunes.
The uncertainty originating from the input PDFs is approximately $\pm5\%$ for both signal and background.
This analysis has several large theoretical uncertainties and studies using early \pp\ data at LHC could provide guidance as to the appropriate choice of \fr\ and \mc\ tune.\\ \\
A light SM Higgs boson will be very challenging to identify at the LHC and several channels will be required to confirm any observation.
Consequently, this analysis provides an important contribution to the overall sensitivity for the $H\to b\bar{b}$ decay mode.
\begin{table*}[htdp] 
\begin{center} 
\begin{tabular}{|l|rr|rr|rr|rr|rr|rr|rr|}\hline \hline 
Cut \# &\multicolumn{2}{c|}{$m_h$ 115 GeV/$c^2$} & \multicolumn{2}{c|}{$b\bar{b}$+2jets+$\gamma$} & \multicolumn{2}{c|}{$b\bar{b}$+1jet+$\gamma$} & \multicolumn{2}{c|}{4jets+$\gamma$} & \multicolumn{2}{c|}{3jets+$\gamma$} & \multicolumn{2}{c|}{z+2jets+$\gamma$} & \multicolumn{2}{c|}{z+1jet+$\gamma$} \\ \hline 
None & \multicolumn{2}{l|}{5096} & \multicolumn{2}{l|}{23503936} & \multicolumn{2}{l|}{41227936} &  \multicolumn{2}{l|}{313253568} & \multicolumn{2}{l|}{815215040}  & \multicolumn{2}{l|}{812711}  & \multicolumn{2}{l|}{1099440} \\ 
1 & 2460  &  $(48\%)$ & 8631124  &  $(37\%)$ & 11964605  &  $(29\%)$ & 116548496 &  $(37\%)$ & 254604944  &  $(31\%)$ & 363997 &  $(45\%)$ & 409825  &  $(37\%)$\\ 
2  & 1835  &  $(75\%)$ & 4744277  &  $(55\%)$ & 622729  &  $(5\%)$ & 80952640 & $(69\%)$ & 19777422 & $(8\%)$ & 238115 & $(65\%)$ & 28732 & $(7\%)$\\ 
3  & 1768 & $(96\%)$ & 4649472 & $(98\%)$ & 608376 & $(98\%)$ & 79378192 & $(98\%)$ & 19324588 & $(98\%)$ & 232430 & $(98\%)$ & 27220 & $(95\%)$\\ 
4   & 438 & $(25\%)$ & 499215 & $(11\%)$ & 48787 & $(8\%)$ & 76163 & $(0\%)$ & 10310 & $(0\%)$ & 7405 & $(3\%)$ & 608 & $(2\%)$\\ 
5 & 336 & $(77\%)$ & 230068 & $(46\%)$ & 9412 & $(19\%)$ & 30275 & $(40\%)$ & 2181 & $(21\%)$ & 4035 & $(54\%)$ & 149 & $(24\%)$\\ 
6 & 221 & $(66\%)$ & 66128 & $(29\%)$ & 1182 & $(13\%)$ & 7766 & $(26\%)$ & 649 & $(30\%)$ & 901 & $(22\%)$ & 13 & $(8\%)$\\ 
7 & 219 & $(99\%)$ & 64744 & $(98\%)$ & 1182 &  & 7707 & $(99\%)$ & 648 &  & 815 & $(90\%)$ & 13 & \\ 
8 & 215 & $(98\%)$ & 60848 & $(94\%)$ & 1115 & $(94\%)$ & 7364 & $(96\%)$ & 583 & $(90\%)$ & 788 & $(97\%)$ & 12  & \\ 
9  & 179 & $(83\%)$ & 36085 & $(59\%)$ & 534 & $(48\%)$ & 3390 & $(46\%)$ & 253 & $(43\%)$ & 667 & $(85\%)$ & 9 & $(68\%)$\\ 
10 & 175 & $(98\%)$ & 34991 & $(97\%)$ & 533  &  & 3066 & $(90\%)$ & 253 &  & 659 & $(99\%)$ & 8 & $(97\%)$\\ 
11 & 129 & $(74\%)$ & 13131 & $(38\%)$ & 123 & $(23\%)$ & 1465 & $(48\%)$ & 65 & $(26\%)$ & 123 & $(19\%)$ & 1 & $(15\%)$\\ 
12 & 110 & $(86\%)$ & 4202 & $(32\%)$ & 26 & $(21\%)$ & 369 & $(25\%)$ & 11 & $(17\%)$ & 72 & $(58\%)$ & 1 & $(95\%)$\\ 
13  & 68 & $(62\%)$ & 1182 & $(28\%)$ & 13 & $(50\%)$ & 110 & $(30\%)$ & 6 & $(49\%)$ & 25 & $(35\%)$ & 1 & $(64\%)$\\ 
\hline \hline
\end{tabular}
\caption{ \label{T:mh115_results}
Number of expected events for 100 fb$^{-1}$ of integrated luminosity for a 115~GeV/c$^{2}$ mass Higgs and background samples at each
step of the selection process for 14 TeV \pp\ collisions. The relative efficiency for each selection step is given in parentheses. The Cut \# is defined in Table \ref{T:cutlist}.} 
\end{center} 
\end{table*}

\begin{table*}[htdp] 
\begin{center} 
\begin{tabular}{|l|rr|rr|rr|rr|rr|rr|rr|}\hline \hline 
Cut \# & \multicolumn{2}{c|}{$m_h$ 125 GeV/$c^2$} &  \multicolumn{2}{c|}{$b\bar{b}$+2jets+$\gamma$} &  \multicolumn{2}{c|}{$b\bar{b}$+1jet+$\gamma$} &  \multicolumn{2}{c|}{4jets+$\gamma$} &  \multicolumn{2}{c|}{3jets+$\gamma$} &  \multicolumn{2}{c|}{z+2jets+$\gamma$} &  \multicolumn{2}{c|}{z+1jet+$\gamma$} \\ \hline 
None  & 4087 &   & 23503936 &   & 41227936 &   & 313253568 &   & 815215040 &   & 812711 &   & 1099440 &  \\ 
1  & 1989 & $(49\%)$ & 8631124 & $(37\%)$ & 11964605 & $(29\%)$ & 116548496 & $(37\%)$ & 254604944 & $(31\%)$ & 363997 & $(45\%)$ & 409825 & $(37\%)$\\ 
2 & 1522 & $(77\%)$ & 4744277 & $(55\%)$ & 622729 & $(5\%)$ & 80952640 & $(69\%)$ & 19777422 & $(8\%)$ & 238115 & $(65\%)$ & 28732 & $(7\%)$\\ 
3  & 1466 & $(96\%)$ & 4649472 & $(98\%)$ & 608376 & $(98\%)$ & 79378192 & $(98\%)$ & 19324588 & $(98\%)$ & 232430 & $(98\%)$ & 27220 & $(95\%)$\\ 
4  & 371 & $(25\%)$ & 498435 & $(11\%)$ & 49093 & $(8\%)$ & 77162 & $(0\%)$ & 10467 & $(0\%)$ & 7427 & $(3\%)$ & 607 & $(2\%)$\\ 
5  & 282 & $(76\%)$ & 229971 & $(46\%)$ & 9396 & $(19\%)$ & 29671 & $(38\%)$ & 3021 & $(29\%)$ & 4021 & $(54\%)$ & 143 & $(24\%)$\\ 
6  & 187 & $(66\%)$ & 65972 & $(29\%)$ & 1400 & $(15\%)$ & 7374 & $(25\%)$ & 849 & $(28\%)$ & 902 & $(22\%)$ & 14 & $(10\%)$\\ 
7  & 186 & $(99\%)$ & 64561 & $(98\%)$ & 1399 &   & 7296 & $(99\%)$ & 848 &   & 822 & $(91\%)$ & 14 &  \\ 
8  & 182 & $(98\%)$ & 60720 & $(94\%)$ & 1304 & $(93\%)$ & 6783 & $(93\%)$ & 836 & $(99\%)$ & 792 & $(96\%)$ & 14 &  \\ 
9 & 148 & $(81\%)$ & 35765 & $(59\%)$ & 711 & $(55\%)$ & 3883 & $(57\%)$ & 446 & $(53\%)$ & 668 & $(84\%)$ & 10 & $(70\%)$\\ 
10  & 144 & $(98\%)$ & 34670 & $(97\%)$ & 710 &   & 3645 & $(94\%)$ & 446 &   & 659 & $(99\%)$ & 9 & $(97\%)$\\ 
 11 & 108 & $(75\%)$ & 11533 & $(33\%)$ & 164 & $(23\%)$ & 1795 & $(49\%)$ & 241 & $(54\%)$ & 84 & $(13\%)$ & 1 & $(8\%)$\\ 
12  & 94 & $(87\%)$ & 4242 & $(37\%)$ & 57 & $(35\%)$ & 506 & $(28\%)$ & 6 & $(2\%)$ & 38 & $(45\%)$ & 1 & $(96\%)$\\ 
13  & 58 & $(62\%)$ & 1207 & $(28\%)$ & 28 & $(48\%)$ & 86 & $(17\%)$ & 6 & $(93\%)$ & 13 & $(35\%)$ & 1 & $(94\%)$\\ 
\hline \hline
\end{tabular}
\caption{\label{T:mh125_results}
Number of expected events for 100 fb$^{-1}$ of integrated luminosity for a 125~GeV/c$^{2}$ mass Higgs and background samples at each
step of the selection process for 14 TeV \pp\ collisions. The relative efficiency for each selection step is given in parentheses. The Cut \# is defined in Table \ref{T:cutlist}.}  
\end{center} 
\end{table*}
\begin{table*}[htdp] 
\begin{center} 
\begin{tabular}{|l|rr|rr|rr|rr|rr|rr|rr|}\hline \hline 
Cut \# & \multicolumn{2}{c|}{$m_{h}$135 GeV} & \multicolumn{2}{c|}{$b\bar{b}$+2jets+$\gamma$} & \multicolumn{2}{c|}{$b\bar{b}$+1jet+$\gamma$} & \multicolumn{2}{c|}{4jets+$\gamma$} & \multicolumn{2}{c|}{3jets+$\gamma$} & \multicolumn{2}{c|}{z+2jets+$\gamma$} &\multicolumn{2}{c|}{z+1jet+$\gamma$} \\ \hline 
None  & 2270  &   & 23503936  &   & 41227936  &   & 313253568  &   & 815215040  &   & 812711  &   & 1099440  &  \\ 
1  & 1116  & $(49\%)$ & 8631124  & $(37\%)$ & 11964605  & $(29\%)$ & 116548496  & $(37\%)$ & 254604944  & $(31\%)$ & 363997  & $(45\%)$ & 409825  & $(37\%)$\\ 
2 & 876  & $(78\%)$ & 4744277  & $(55\%)$ & 622729  & $(5\%)$ & 80952640  & $(69\%)$ & 19777422  & $(8\%)$ & 238115  & $(65\%)$ & 28732  & $(7\%)$\\ 
3  & 846  & $(97\%)$ & 4649472  & $(98\%)$ & 608376  & $(98\%)$ & 79378192  & $(98\%)$ & 19324588  & $(98\%)$ & 232430  & $(98\%)$ & 27220  & $(95\%)$\\ 
4  & 216  & $(26\%)$ & 498158  & $(11\%)$ & 49244  & $(8\%)$ & 76238  & $(0\%)$ & 9508  & $(0\%)$ & 7350  & $(3\%)$ & 606  & $(2\%)$\\ 
5  & 166  & $(77\%)$ & 229746  & $(46\%)$ & 9634  & $(20\%)$ & 28590  & $(38\%)$ & 1930  & $(20\%)$ & 4025  & $(55\%)$ & 144  & $(24\%)$\\ 
6  & 111  & $(67\%)$ & 66282  & $(29\%)$ & 1333  & $(14\%)$ & 6878  & $(24\%)$ & 387  & $(20\%)$ & 890  & $(22\%)$ & 12  & $(8\%)$\\ 
7  & 111  & $(99\%)$ & 64810  & $(98\%)$ & 1333  &   & 6844  &   & 386  &   & 804  & $(90\%)$ & 12  &  \\ 
8  & 107  & $(97\%)$ & 61105  & $(94\%)$ & 1271  & $(95\%)$ & 6391  & $(93\%)$ & 372  & $(96\%)$ & 773  & $(96\%)$ & 12  &  \\ 
9  & 86  & $(80\%)$ & 35869  & $(59\%)$ & 598  & $(47\%)$ & 3431  & $(54\%)$ & 172  & $(46\%)$ & 658  & $(85\%)$ & 9  & $(76\%)$\\ 
10  & 84  & $(98\%)$ & 34714  & $(97\%)$ & 596  &   & 3291  & $(96\%)$ & 169  & $(99\%)$ & 648  & $(98\%)$ & 9  & $(97\%)$\\ 
11  & 62  & $(74\%)$ & 10053  & $(29\%)$ & 118  & $(20\%)$ & 1405  & $(43\%)$ & 23  & $(14\%)$ & 69  & $(11\%)$ & 0  & $(1\%)$\\ 
12  & 54  & $(87\%)$ & 3728  & $(37\%)$ & 40  & $(34\%)$ & 549  & $(39\%)$ & 7  & $(29\%)$ & 27  & $(39\%)$ & 0  & $(87\%)$\\ 
13  & 33  & $(61\%)$ & 1091  & $(29\%)$ & 27  & $(69\%)$ & 76  & $(14\%)$ & 5  & $(74\%)$ & 11  & $(41\%)$ & 0  & $(20\%)$\\ 
\hline \hline
\end{tabular}
\caption{\label{T:mh135_results}
Number of expected events for 100 fb$^{-1}$ of integrated luminosity for a 135~GeV/c$^{2}$ mass Higgs and background samples at each
step of the selection process for 14 TeV \pp\ collisions. The relative efficiency for each selection step is given in parentheses. The Cut \# is defined in Table \ref{T:cutlist}.} 
\end{center} 
\end{table*}

\section{Acknowledgments}\label{S:acknow}
We wish to thank Aleandro Nisati, Barbara Mele and Fulvio Piccinini for the discussions which initiated this analysis.
Additionally, we wish to thank Michelangelo Mangano (\alpgen) and Torbjorn Sjostrand (\pythia) for providing us guidance, advise, and support  regarding the use of their \mc\ generators.
This research was funded by Natural Sciences and Engineering Research Council of Canada, Fonds de recherche sur la nature et les technologies, and Canada Research Chairs.
\newpage

\end{document}